RADIOCOMMUNICATIONS AGENCY

RCRU

# RADIO AND THE 1999 UK TOTAL SOLAR ECLIPSE

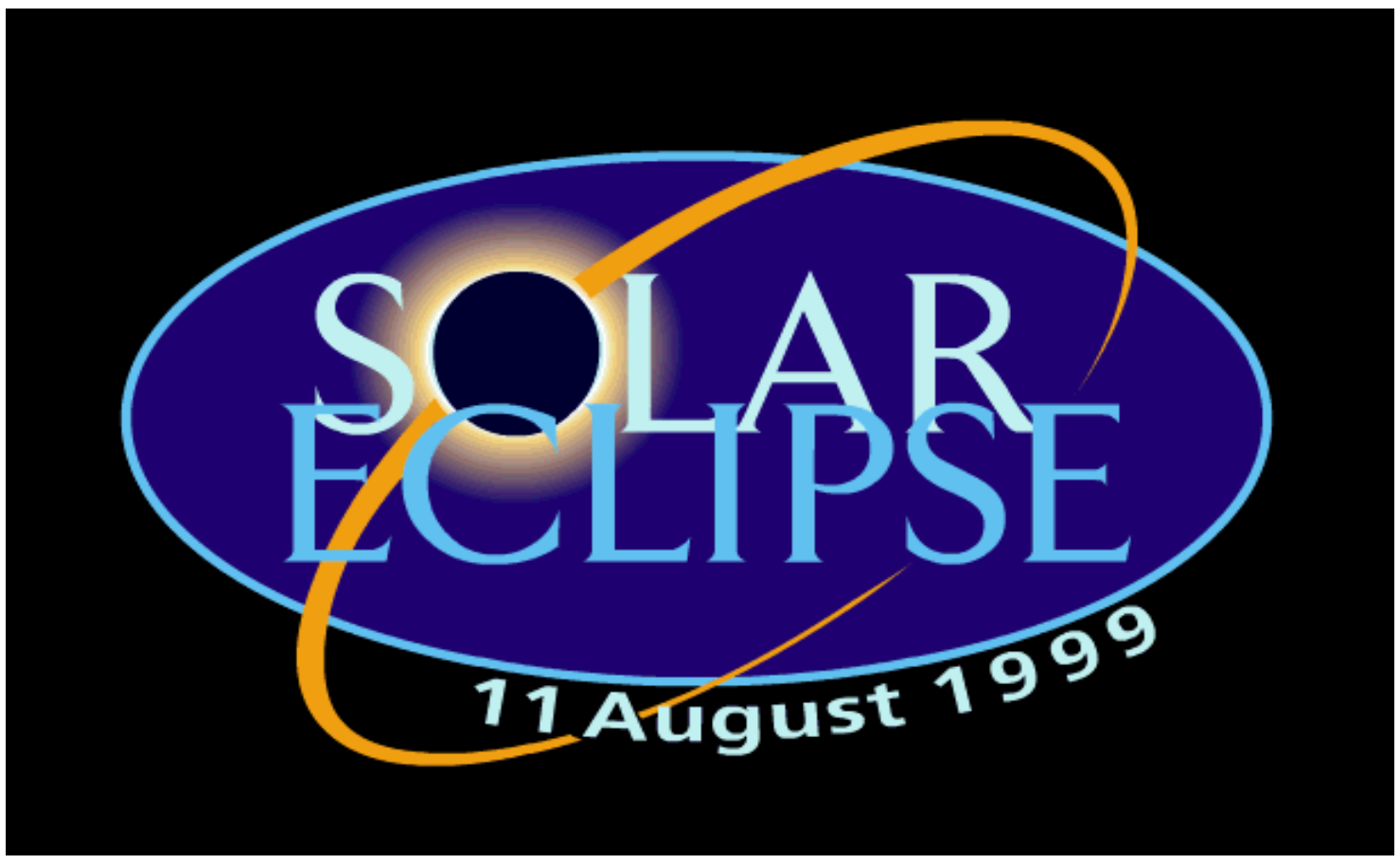

Project Final Report

May 2000

**Dr. Ruth Bamford**

*Radio Communication Research Unit,*
*Rutherford Appleton Laboratory,*
*Chilton, Didcot, OXON OX11 0QX, UK*

*Tel. 01235 44 6517, E-mail: R.Bamford@rl.ac.uk*

# *ABSTRACT*

*On the morning of the August 11th 1999, a total eclipse of the sun plunged Cornwall and parts of Devon into darkness. The event of the eclipse was bound to attract a great deal of scientific and media attention. Realizing that the differences in day-time/night-time propagation of VLF/LF/MF to HF bands would also apply during the darkness of the eclipse, the eclipse offered a rare PR opportunity to promote radio to the general public. At the same time the specific nature of the disturbance to the upper atmosphere and the effect on radio propagation could be examined in detail using scientific instruments at minimum cost since most instruments would not have to be moved. This would allow prediction models to be tested in a controlled fashion.*

*Contained within this report are the details and results of the radio and ionospheric experiments conducted by the Rutherford Appleton Laboratory during the 1999 total solar eclipse. The promoting of the radio experiments with the general public produced nearly 60 appearances on local and national TV, newspapers and periodicals. Close to 1700 people responded to the general public medium wave experiment and 16 million people looked in on the general eclipse web site (part funded by RA) that included the details of the radio experiments. A large database of systematic observations across VLF to HF was collected from radio amateurs and from the RA Regional Offices allowing comparisons to be made with ITU estimates. There is a brief look at the scientific results and a forward look as to how the analysis of this disturbance might have impact on the use of ionospheric models for Space Weather tools in the future.*

# ACKNOWLEDGEMENTS

This project would not have been possible if it were not for the help of many people. Most particularly for Dr. Chris Davis of the Space Science and Press & PR Departments of the Rutherford Appleton Laboratory who originally pointed out that the eclipse could be used as an ionospheric experiment and worked very hard on the PR for the radio experiments along with the ionospheric experiment. Thanks to David Eden of RA for supporting this project from the beginning without which it would never have happened either. To Richard Stamper from the World Data Center for running and maintaining the Website despite incredible hit rate on the morning of the eclipse. To Jim Finnie, Ray Wiltshire and Tony Beard, Richard Drinkwater of RA for their support and active involvement. To Lance Farr my undergraduate student for the year for typing in and helping to analyse the radio amateur data and produce plots. I would also like to thank Mike Willis of the RCRU and Pete Mahy without whose help the radio scanner experiment would not have been possible. Thanks also to Graham Burt of Digital Artwork for producing the radio graphics and animations as well as the eclipse logo. Thanks to Fred Espenak of NASA, EUMETSAT, M. Sanders, Matthew Harris of UCL, members of the Southampton Uni. for the use of their images within this report. Thanks is also due to PPARC the Particle Physics and Astronomy Research Council for their funding of the ionosonde project during the eclipse. And a special thanks to Gill Spong of the RCRU for her help in creating the certificates. And last but by no means least at deal of thanks goes to all those members of the Radio Society of Great Britain (RSGB) and the general public who took the trouble to do the eclipse radio experiments and post their results to the Rutherford Appleton Laboratory.

# TABLE OF CONTENTS

# LIST OF FIGURES

# INTRODUCTION

On the August 11th 1999, for the first time in 72 years, a total solar eclipse of the sun was visible on the UK mainland. Although the region of 100% eclipse was mainly over Cornwall, the partial shadow extended across the whole of the UK, with as much as 85% of the sun's disk being obscured by the moon as far north as Edinburgh. This event was bound to create a great deal of public, scientific and media interest.

It is not immediately apparent that a total solar eclipse would have anything to do with radio propagation. However it is well know that Medium Wave, Long Wave and Short Wave reception can be quite different at night compared to during the day. During the night many remote MW and LW stations are detectable on the band that are not audible during the day. This is due to changes in the ionosphere, the layer of the Earth’s upper atmosphere that is ionised into free electrons and ions by radiation coming from the sun. Since this layer of the atmosphere is comprised of electrical charges it can effect radio waves. Also the nature of the ionosphere is so closely tied to radiation from the sun, the loss of sunlight due to the passage of the shadow of the moon during the eclipse was going to very briefly produce a night-time like ionosphere. In essence this meant that radio reception on the Medium, Long and Short Wave bands were expected to experience the sudden and relatively localised return to night-time reception conditions during the middle of the day over Europe on August 11th 1999. However the darkness that occurs during an eclipse is considerably different for that of an ordinary night-time. The moon’s shadow is relatively small on the Earth and travels at super sonic speeds. It was likely to produce some interesting effects that might be detectable on ordinary radios.

This event offered a two-fold opportunity. Firstly because the eclipse was going to effect household medium wave (MW) radio reception (though not FM), it was a potential public involvement event. This was going to make it a rare opportunity to put radio and the activities of the Radio communications Agency into the media. Secondly it was an opportunity to study the effects on radio propagation of the sun through its effects of the ionosphere. The eclipse was particularly useful for this because the nature of the eclipse disturbance could be exactly characterised/predicted which is not so true of events like solar flares or solar storms that more commonly disturb ionospheric sensitive communications and navigation systems such as GPS (Global Positioning System).

# PROJECT OBJECTIVES

The main objectives of the project are as follows:

(i) Increase public awareness of the effects of sun and night-time conditions on radio reception in both the medium and short wave bands.

(ii) Promote better links between propagation researchers and the short wave radio user community by working together on a specific event.

(iii) Use the unique circumstance of a total solar eclipse to examine ionospheric models.

This report describes both the public and scientific radio propagation programs conducted during the total solar eclipse over the UK and Europe in August 1999. A summary of the results is included and a breakdown of the media and web outlets used. The introductionary sections however start with a description about the nature of eclipses and the principles behind the effects on the ionosphere and radio propagation.

# THE NATURE OF AN ECLIPSES

In Figure 1 is the familiar image of a total solar eclipse with the moon obscuring the face of the sun leaving just the corona. Remarkably, the fact that the moon's diameter so exactly matches that of the sun's face viewed from Earth, is just a co-incidence. The moon is 400 times smaller in diameter than the sun but is 400 times closer (see Figure 2).

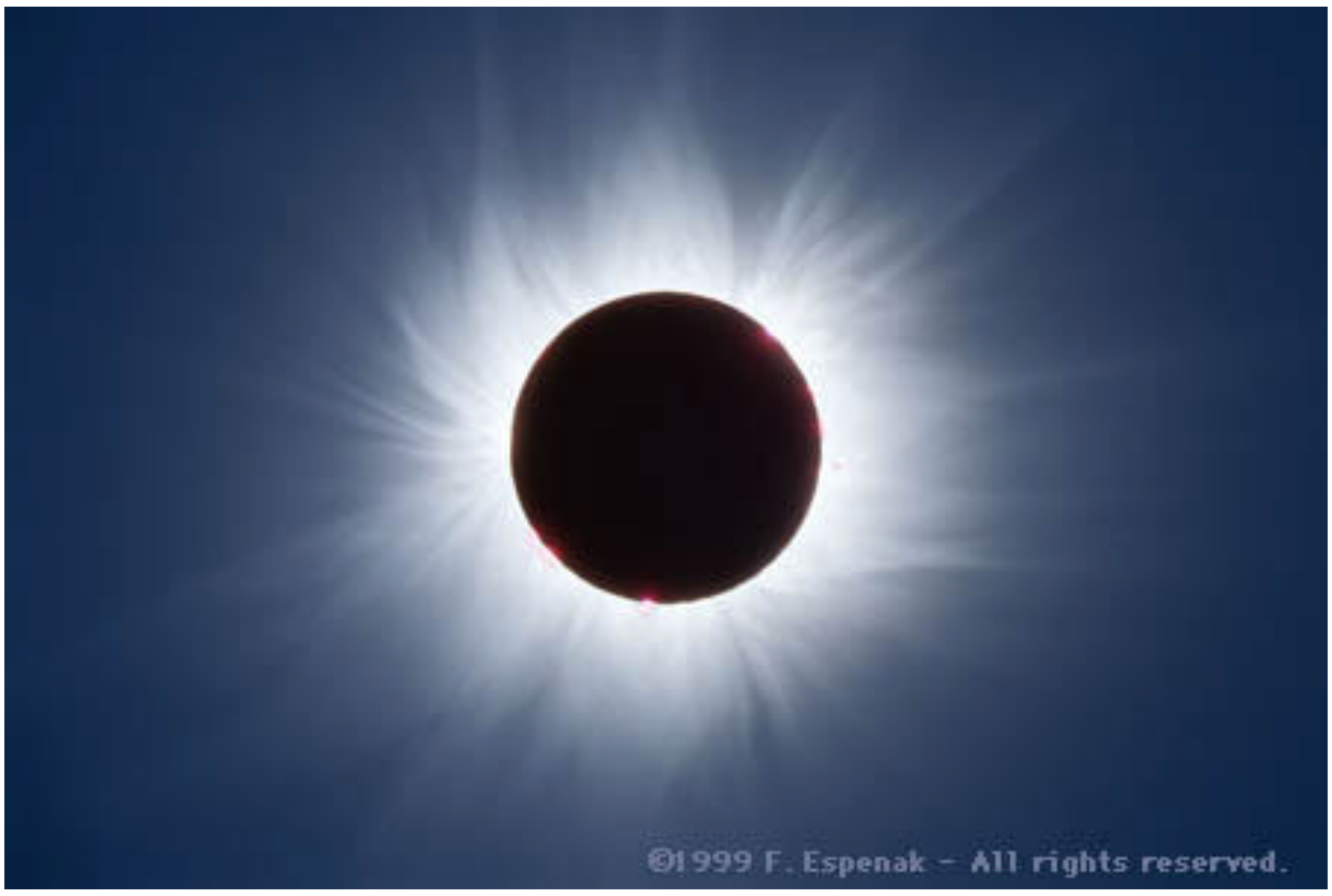

*Figure 1. A photograph of the Total Eclipse of the Sun in August 11th 1999.*

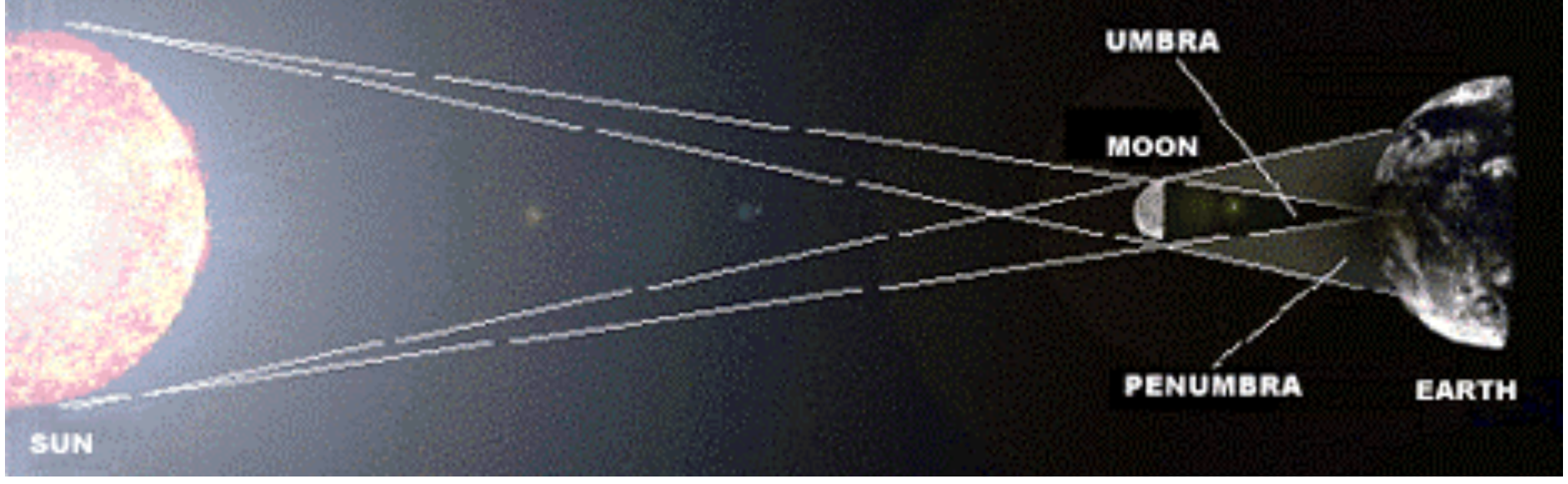

*Figure 2. The geometry of an eclipse showing the regions of 100% shadow (umbra) and partial shadow (preumbra). (Image from F. Espenak, NASA).*

On average there is either a total or annular eclipse about twice a year somewhere on the Earth. But the Earth is big and the shadow of the moon during an eclipse is very small by comparison making it a rare event for the UK. The last total solar eclipse on the UK mainland was in 1927 and the next will not be until 2090.

The photograph of the 1999 eclipse taken by a weather satellite METEOSAT-7 in Figure 3 best illustrates the size of the moon's shadow on the earth during an eclipse. In this image the 100% eclipse, or totality region, is centred over Northern France and is only 100km wide. However the partial shadow can be seen to extend over most of Europe. In motion, the eclipse shadow passes across this image from left to right in about an hour and a half, which is at super sonic speed.

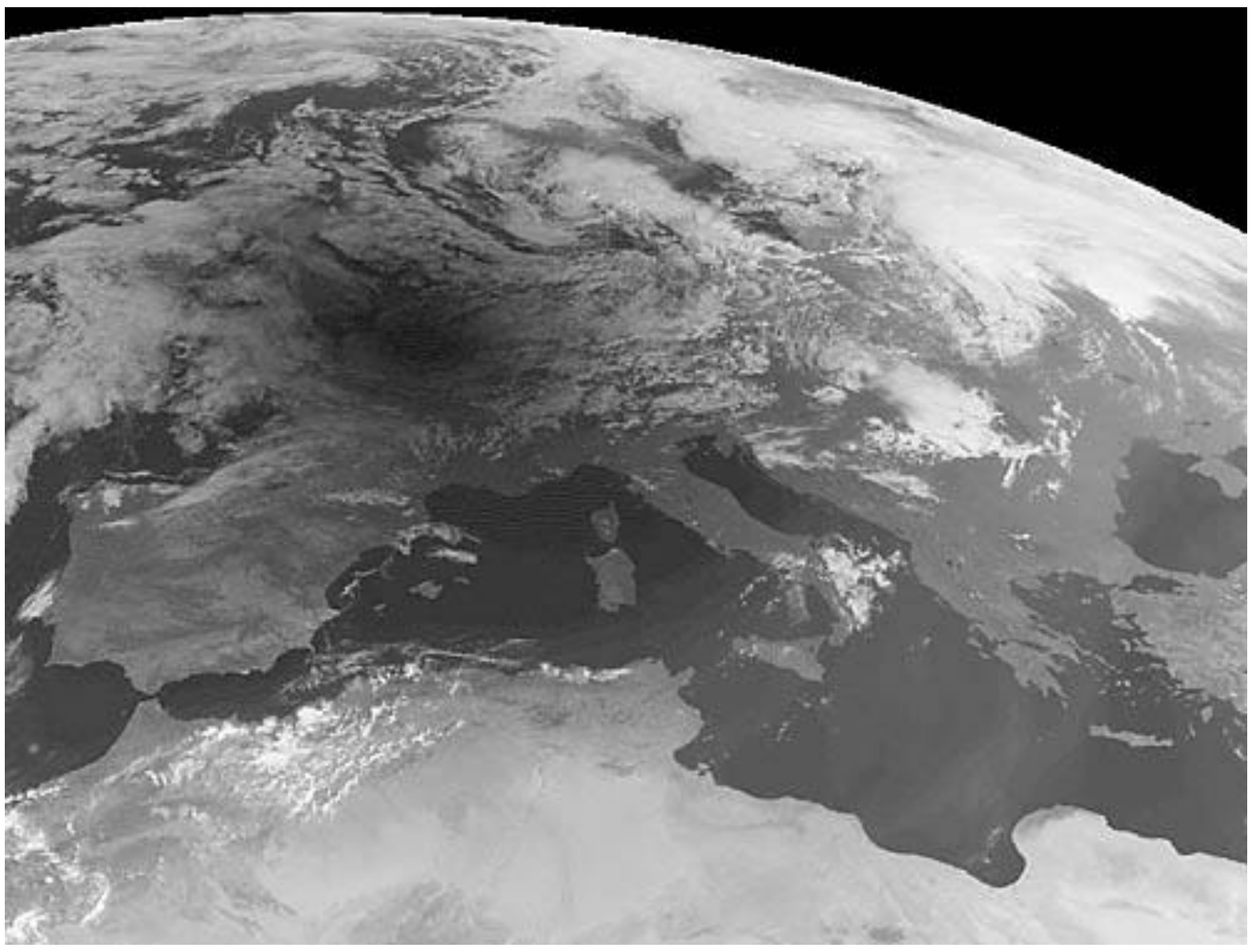

*Figure 3. Topside view. A photograph of the 1999 total solar eclipse at 10:30UT taken by a weather satellite (Meteosat-7). © 2000 EUMETSAT.*

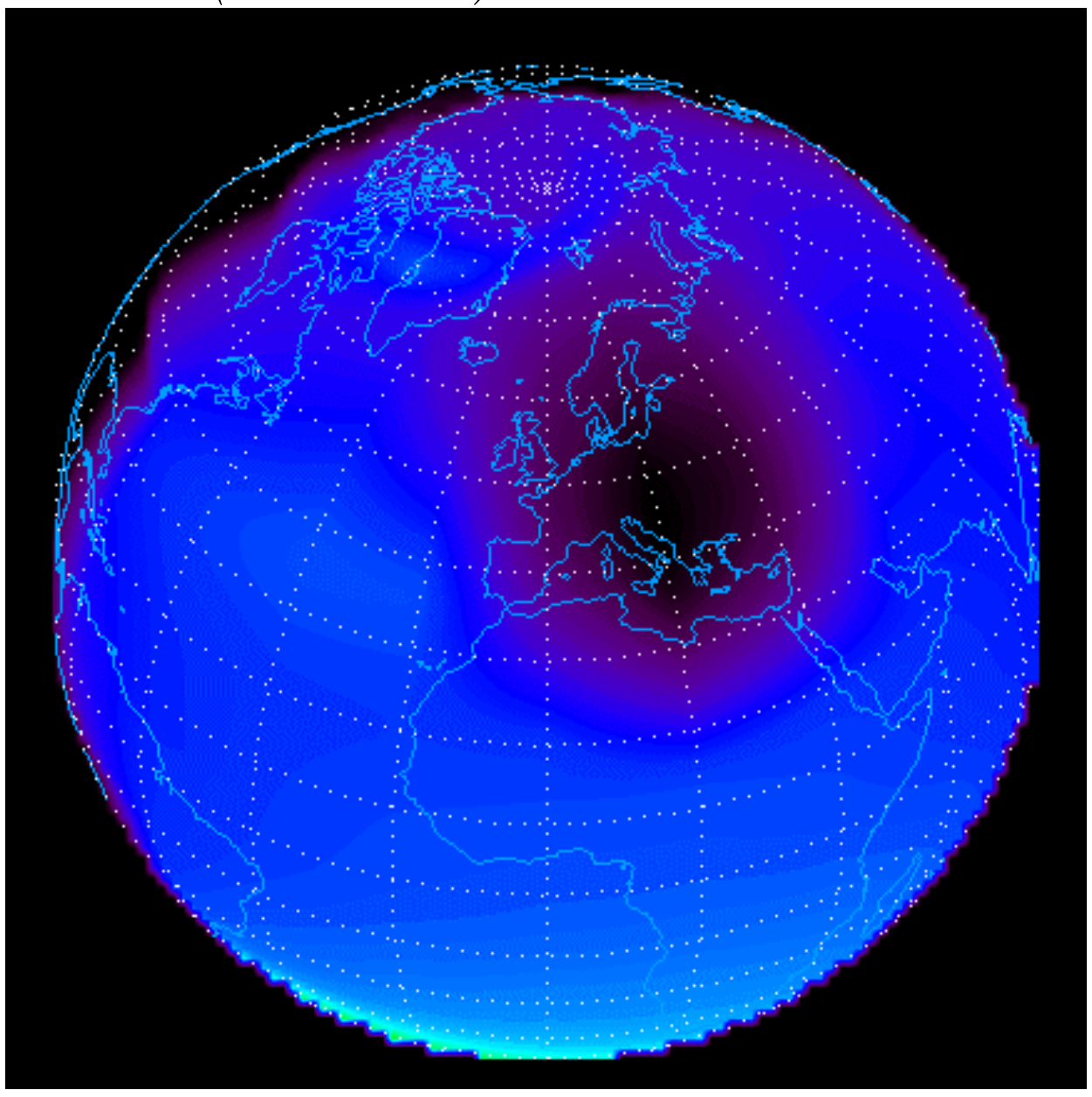

*Figure 4. A model calculation of the depletion in electrons caused by the lunar shadow on the Earth's ionosphere. (Matthew Harris, University College London).*

The image in Figure 4 shows the calculated effect of the lunar shadow on the ionosphere, the ionised upper most part of the Earth's atmosphere that effects radio propagation. The colours represent the calculated electron density in the ionosphere at an altitude of ~280km. Since the electrons in the ionosphere are created by the photoionisation of the sun's radiation, the simulation shows that as expected, without the sunshine and electron density drops, giving us a nice outline of the lunar shadow very similar to the METEOSAT image in Figure 3. This illustrates how the well the dimensions and motion of this disturbance to the ionosphere can be modelled and thus used to test ionospheric and radio propagation predictions effected by the ionosphere.

Figure 5 shows where these previous two figures relate a map of Europe. The time at which the shadow passed is indicated along the path as is the geographical extent of the partial shadow.

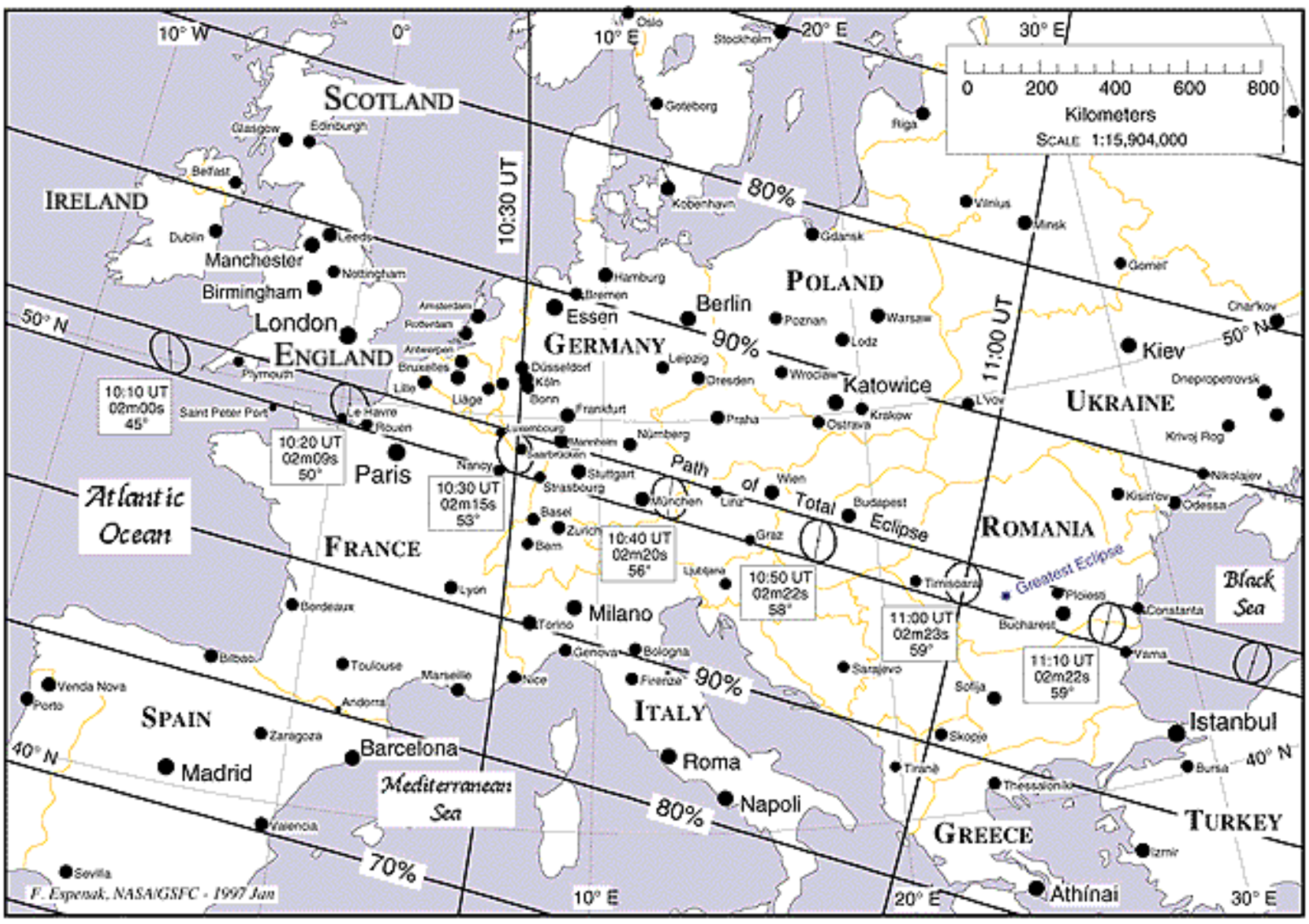

*Figure 5. A map of the path of totality across Europe of the 11th August 1999 total solar eclipse. (From F. Espenak, NASA).*

# WHY THE ECLIPSE WAS ABLE TO EFFECT RADIO PROPAGATION

The connection between the solar eclipse and radio propagation is due to the connection with ELF to HF radio bands and the ionosphere. Therefore it is necessary to give a brief introduction to the nature of the ionosphere and how the loss of sunlight from an eclipse effects the ionosphere before discussing how the radio experiments during the eclipse worked.

## *THE IONOSPHERE*

The ionosphere extends from about 50km altitude to about 600km where it merges with near Earth space environment. The ionosphere is created by radiation, both electromagnetic and particles, coming from the sun which ionise the Earth's upper atmosphere into free electrons and ions. The reason the free electrons persist in the ionosphere is that at these altitudes the density of the Earth's neutral atmosphere is sufficiently thin that collisions between particles happen far less frequently than in the lower atmosphere. However even at these altitudes the particles of the neutral atmosphere within the ionosphere still out number the free electrons by about 10,000:1. But even so the free electrons live much longer before they are recombined.

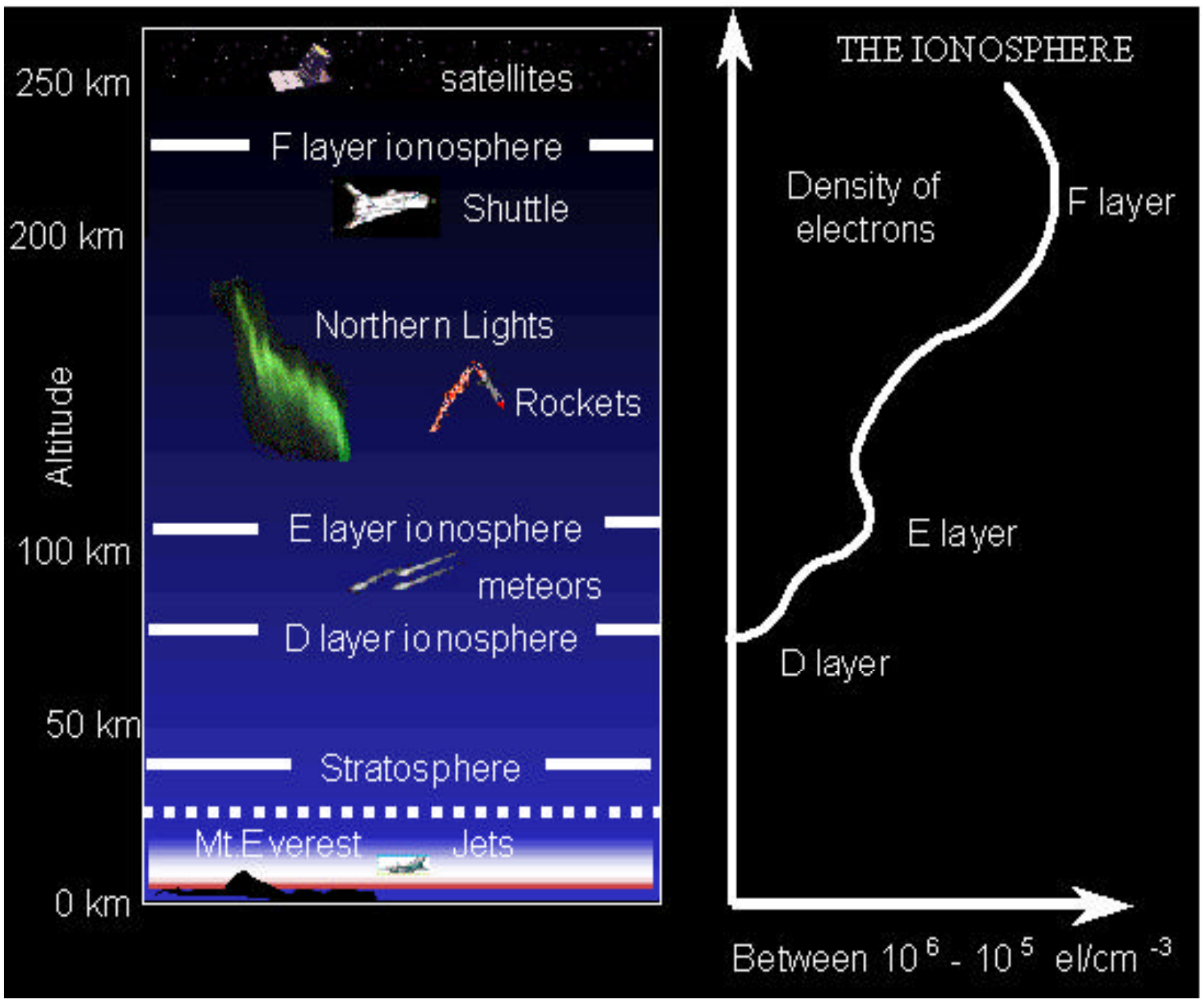

*Figure 6. The location of the layers of the ionosphere in altitude and a sketch showing the peaks in the electron concentration that are referred to as the D, E and F layers of the ionosphere.*

The density or concentration of free electrons varies greatly with altitude within the ionosphere. This is in accordance with the balance between the production (from the sun mainly) and loss mechanisms (through collisions). As a result of the variations in the composition of the Earth's neutral atmosphere and the different penetration depths of the solar emissions, the ionosphere electron density profile has a complicated anatomy that is generally divided into several regions or layers (see Figure 6).

The regions of the ionosphere are labelled the D, E and F layers for historical reasons. The F layer is the highest layer where the atmosphere is the thinnest. It also has the maximum density of electrons overall and ground based short-wave or High Frequency (HF 3-30MHz) radio signals generally use this layer to reflect from during the day. The D layer of the ionosphere is the lowermost and is most dominated by in the neutral atmosphere. The Extremely Low, Very Low and Low Frequency bands (ELF, VLF and LF, <3kHz to 300kHz) can reflect off the D layer of the ionosphere. The E layer is the layer in between these two and the Low, Medium and High Frequencies (LF, MF and HF 30kHz to 30MHz) can reflect from this layer. Because the ionosphere changes with the variations in solar emission and changes in the "weather" of the neutral upper atmosphere, the efficiency of the propagation and signal reception of all these bands therefore depends heavily on the state of the ionosphere. So if the ionosphere is disturbed so will radio propagation at ELF to HF frequencies.

## *THE DIFFERENCES IN RADIO RECEPTION WITH AND WITHOUT THE SUN*

Each of the layers of the ionosphere respond differently to the loss of sun's radiation at night or during an eclipse. The D layer of the ionosphere is the layer responsible for most of the *signal absorption*. It is also the most sensitive to the loss of sunlight. This is because it is the lowermost of the layers and is quickly overwhelmed by the neutral air around it once the active source ionizing radiation from the sun is removed.

However the E and F layers above the D layer are more resilient to the loss of solar radiation. The electron densities are much greater in these layers and they are higher up in the Earth's atmosphere too so that they persist for much longer, even for a while after sunset.

The sketch in Figure 7 shows how the D layer disappears during the eclipse but the layers above may remain for longer. In the case of an eclipse the "hole" in the D layer of the ionosphere is a much more localized effect than during a normal night.

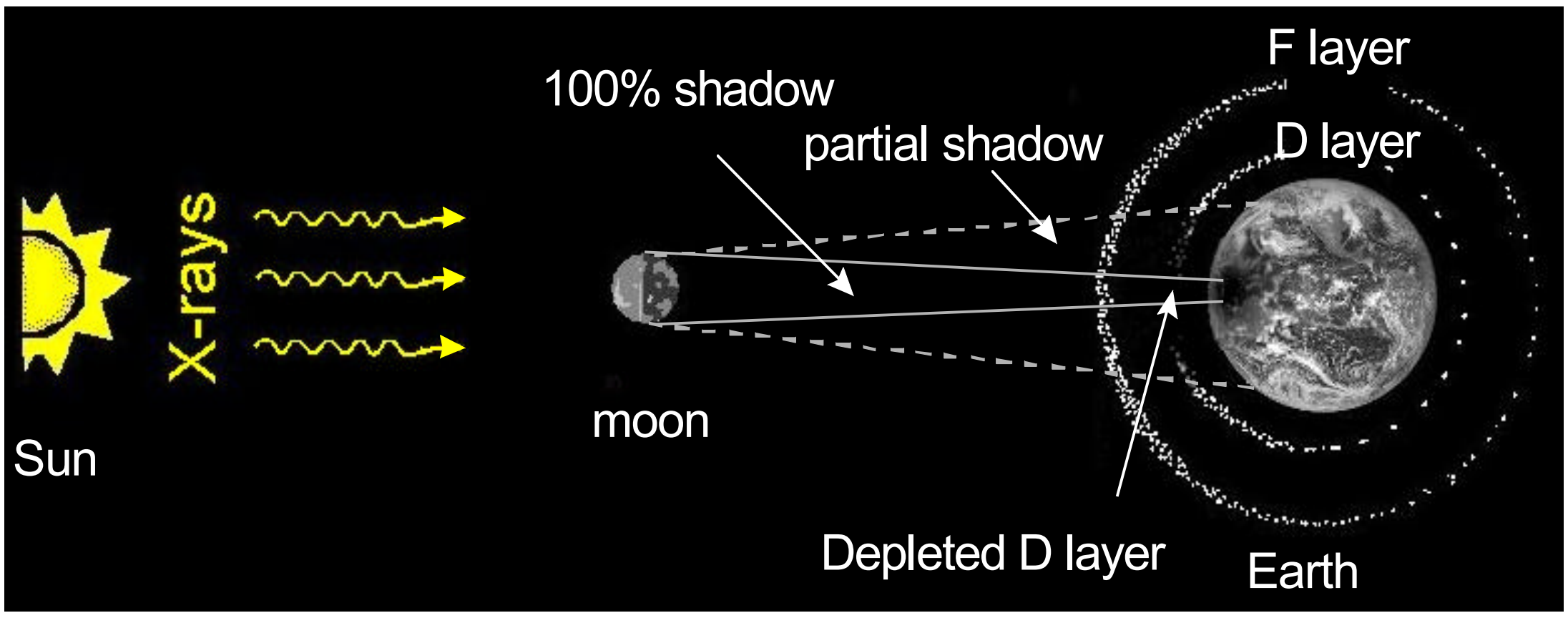

*Figure 7. Radiation from the sun is what largely creates the ionosphere. The shadow cast by the moon onto the Earth puts the lowermost part of the ionosphere, the D layer, into its night-time like state. The E & F layers are more resistant to change*

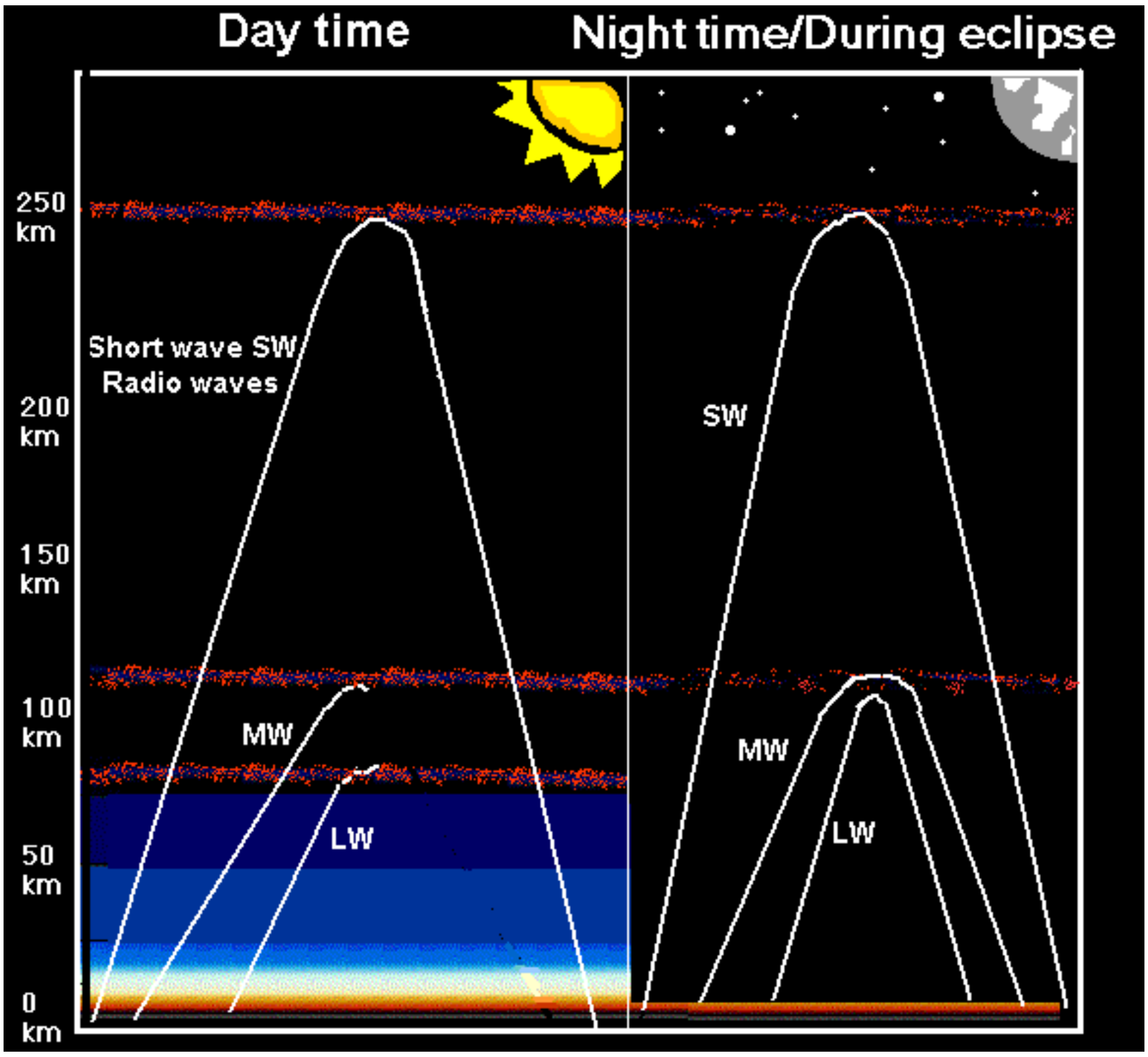

*Figure 8. A sketch illustrating how long wave (LW, 30-300kHz LF), medium wave (MW, 300kHz to 3MHz MF) and short wave (SW, 3-30MHz HF) radio waves are reflected from the different layers of the ionosphere during the day (a) and during an eclipse or at night (b).*

The consequence of the effect of the loss of sunlight either at night or during an eclipse is illustrated in Figure 8.

During the day the emission from the sun means that the electron density of the radio absorbing D layer of the ionosphere is high. For MF and LF band this means that signals from many remote broadcast stations are not detectable during the day, as the signal strength is too low for most receivers and certainly so for most household radios. But at night with the loss of the direct solar radiation the D layer disappears rapidly. However the higher altitude E and F layers remain all be it in a less intense form than during the day. Therefore MF and LF can propagate by reflecting from these layers without the hindrance of the absorption from the D layer. This results in the night-time band pollution from a multitude of foreign stations.

## *ARE THE DIFFERENCES SIGNIFICANT ENOUGH?*

It was important to know at the outset of this project whether or not the effect on the signal strength of sky wave VLF/LF/MF and HF would be of measurable magnitude during an eclipse. The evidence came from the ITU recommendations and reports [ITU 1986] that show typical daytime/night time values for the changes in signal strength. These are shown in Table 1. Since the same processes are at work the moon's shadow would produce similar magnitudes of changes in signal strength but just over a more limited geographical extent. At 10 to 65dB the passage of the eclipse would be easily be detected on normal radio.

*Table 1. Examples of some typical changes in signal strength between normal daytime and night-time.*

| Band | Data for frequencies | Ground ranges | Sig. change |
|---|---|---|---|
| LF | 227kHz - 251kHz | 1380 -1650km | 10 - 20 dB |
| MF | 845kHz - 1538kHz | 610 - 405km | 45 - 65dB. |
| HF | 4.8MHz -15.3MHz | 300 - 2300km | 10 - 30dB |

## *THE RELEVANCE OF THE ECLIPSE TO THE EVERYDAY IONOSPHERE*

Although an eclipse in any one location is a very rare event, observations of the consequences of an eclipse on radio propagation do have a direct connection with more everyday ionospheric disturbances. The effect of the eclipse on the ionosphere can be considered as the opposite a short wave fade out. This is where a solar flare erupts on the sun and emits a burst of x-rays and extreme ultraviolet radiation. This radiation enhances the ionospheric D layer that then increases the amount of RF absorption from VLF to HF disrupting communication on a wide range of frequencies. A sketch illustrating the comparison between a short wave fade out and the eclipse is shown in Figure 9. This figure also demonstrates the need for observations to be made across many frequencies.

The significance of this comparison between a solar flare and the eclipse also illustrates the relationship between the extreme ultraviolet (EUV) wavelengths (below about 20 nm) and X-rays emissions from the sun and the formation of the ionospheric layers. Thus the eclipse offers an opportunity to examine the effect of *regions of the sun* on the ionosphere as the moon progressively obscures the solar disk. This analysis is shown in a later section.

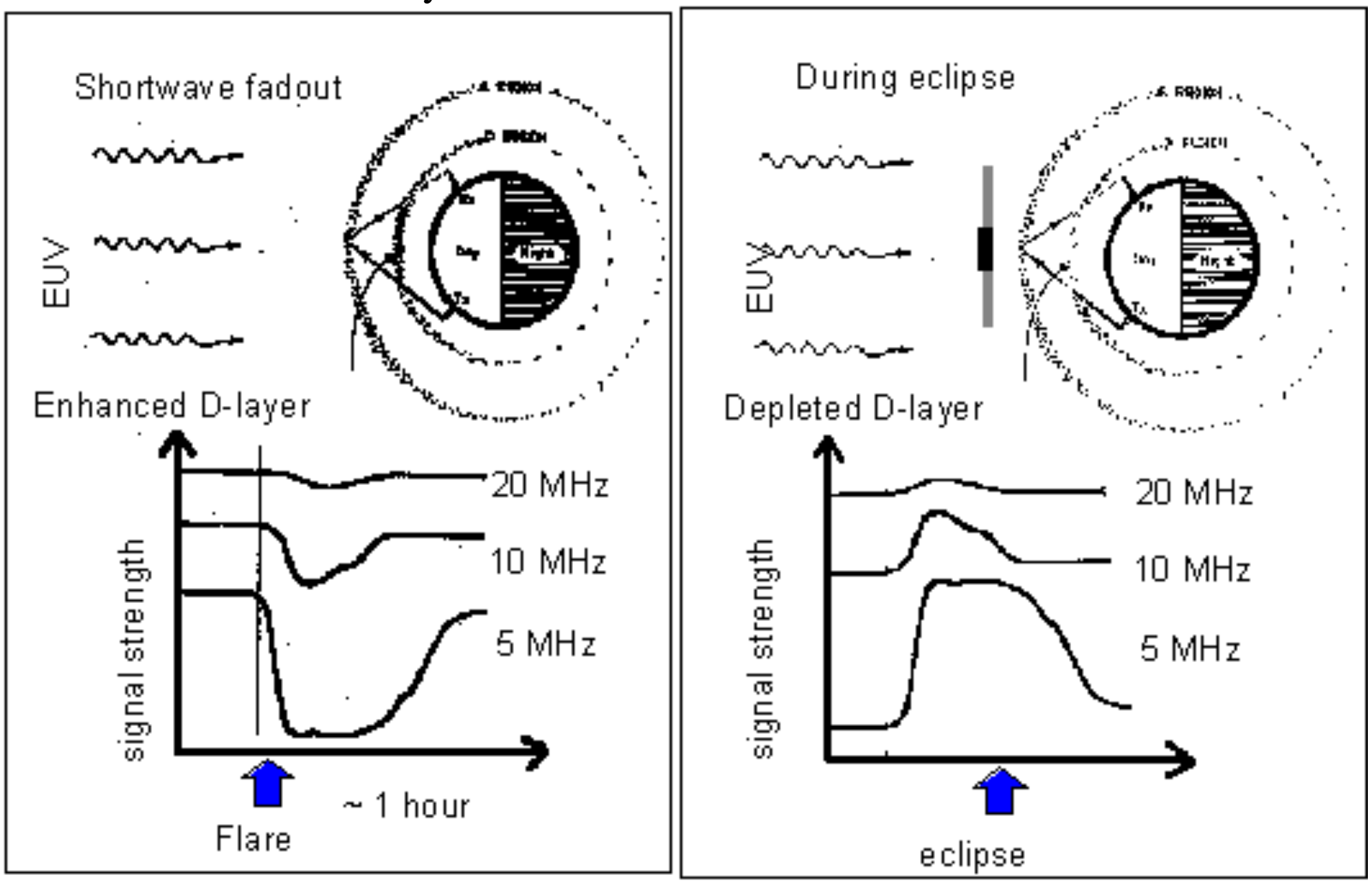

*Figure 9. An illustration of how the eclipse shadow passing over the face of the sun could be expected to enhance signal strengths at different frequencies in the opposite manner to a short wave fade-out created by solar flares.*

# RADIO EXPERIMENTS DURING THE TOTAL SOLAR ECLIPSE

The radio experiments developed for the eclipse came on several levels. Firstly was the general public "did you hear the eclipse? Yes or no?" experiment on the medium wave using household radios. Secondly was the more methodical observations of the radio amateurs from across Europe and the RA regional offices both using more sophisticated radio receivers. And finally were the scientific observations of the ionosphere directly using the HF radars or "ionosondes".

## GENERAL PUBLIC MEDIUM WAVE EXPERIMENT

The loss of the D layer signal absorption during the eclipse meant that for a brief time during the middle of the morning during the eclipse, Continental radio stations not normally audible would appear to *fade in* and *fade out* as the eclipse passed by.

The experiment consisted of asking the general public to tune in their radios to a specific foreign MW radio station during the eclipse and tell us if they could hear it "yes or no". They were also asked whether they could normally hear it during the day, yes or no, and where they were in the country. The station chosen was Radio La Coruna on 629kHz transmitting with a 2KW transmitter from Northern Spain. The station was chosen because it was easily detectable during the night and the frequency was free from any local UK broadcast stations.

The public was asked to send in their responses, with their postcode so that their location could be determined, to the Rutherford Appleton Laboratory. It was as important to receive responses from those who *did not* hear Radio La Coruna during the eclipse as from those that *did* and to know where each were in the country.

The radio experiments had several benefits over other eclipse experiments that added to it success. The experiment was suitable for people living throughout the country, not just those in Cornwall, it involved any age group and included the visually impaired. Two other important factors were emphasised with the media that helped to increase the number of people who participated in the experiment; it was unaffected by cloudy skies and rain and it did not prevent people from watching the eclipse at the same time.

Even if people did not try the experiment themselves the intention was increased public awareness of how radio signals reach them and what factors effect the quality of reception.

An example of some of the professional graphics commissioned to promote the radio experiments is shown in Figure 10. These images helped to visualise the radio experiments to the media. Stills and the animation versions were used by several of the media outlets.

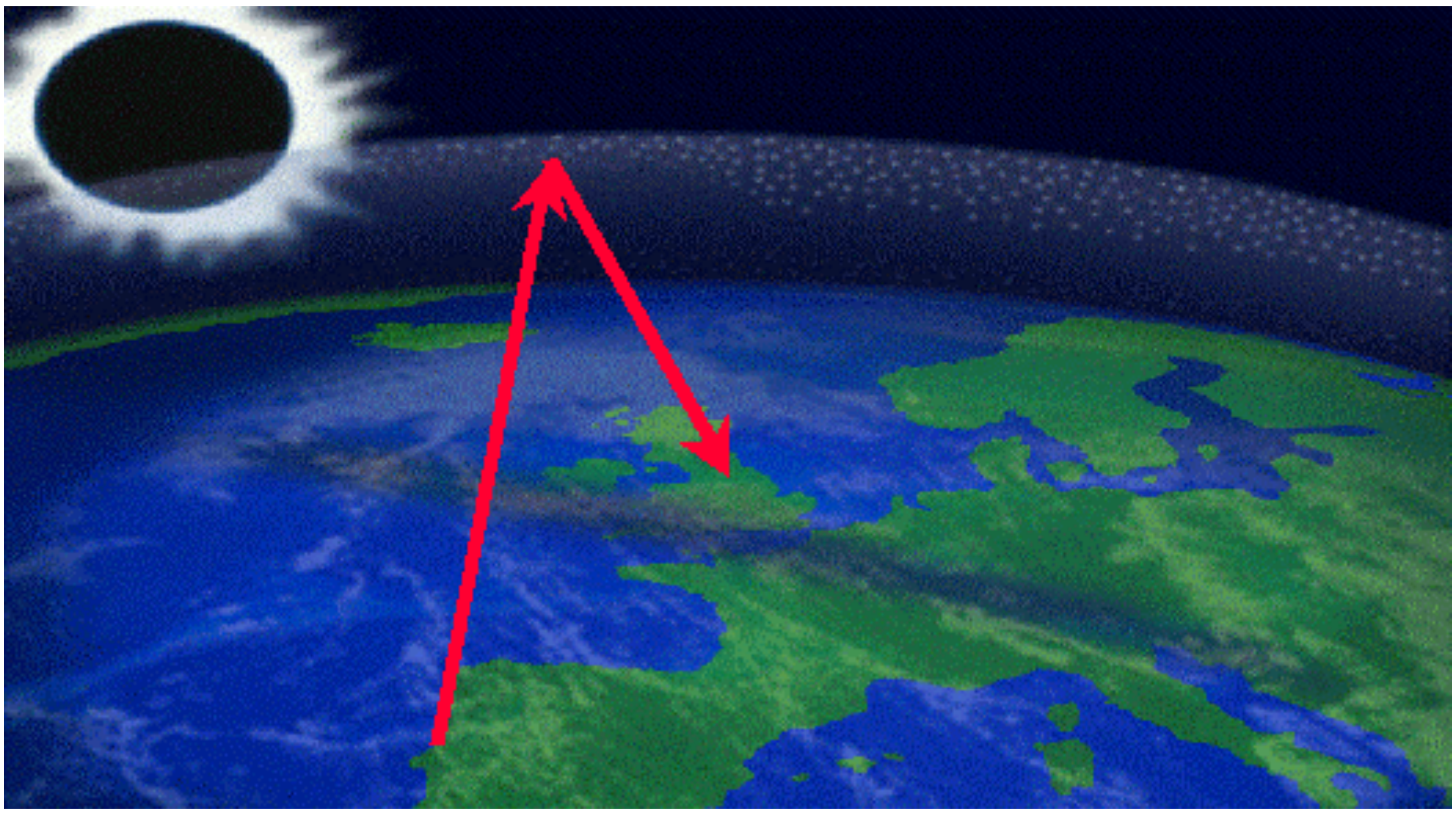

*Figure 10. An example of the graphics used to promote the radio experiment with the general public. The animated version was used by Sky News.*

# RESULTS FROM THE GENERAL PUBLIC RADIO EXPERIMENT

In total 1700 responses were received mainly by post. The size of this response is thanks to the success of the www.eclipse.org.uk website, BBC Online and the Daily Telegraph publishing a coupon on the day of the eclipse for people to fill in and post to the Rutherford Appleton Laboratory (Daily Telegraph 11th August 1999). With this quantity of respondents a map of the country showing where 639kHz La Coruna was audible during the eclipse could be drawn and this is shown in Figure 11. The lighter colour indicates the higher percentage of “yes” responses. The responses across the country were divided into percentage of positive responses to total responses in grid boxes across the UK. This contour map was then extrapolated parallel to the path of the eclipse to produce the image shown in Figure 11. The follow-up article in the Daily Telegraph on Wednesday, September 15th 1999 reproduced this picture.

From the map it can be seen that people in the South and Southwest of the UK could hear 639kHz Radio La Coruna all the time. This was known and is probably due to propagation directly across the sea from Northern Spain. There is a region across Wales and the middle of England where the station was not audible but people further north in the Northeast and Northwest of England the Radio La Coruna was heard during the eclipse. Thus suggesting a skip-distance effect.

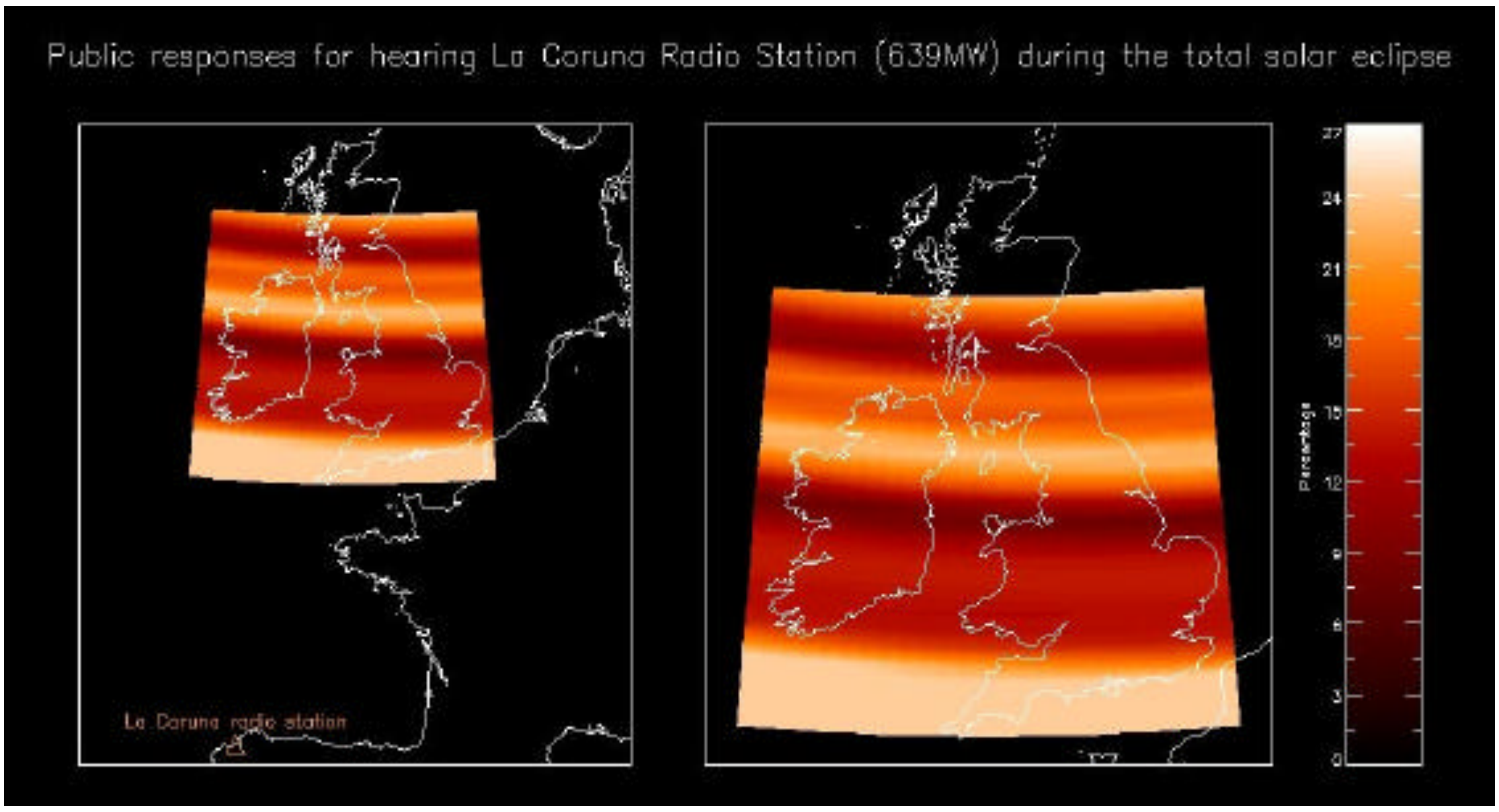

*Figure 11. The distribution of % positive responses to the 639kHz Radio La Couruna public experiment. (Thanks to Lance Farr).*

# THE VLF TO HF BANDS

### *RADIO AMATEUR EXPERIMENTS*

The radio amateurs with their more sophisticated radio equipment were in a position to make more systematic observations and cover a much wider band of frequencies. This event was also an opportunity to involve radio amateurs directly with RA and radio researchers at RAL. Scientifically it provided an opportunity to compare the observations during an eclipse with those published by the International Telecommunications Union (ITU) for day-time/night-time differences and to model the consequences of the effects on the ionosphere.

The experiment for the radio amateurs and RA regional offices was essentially a classical A3 ionospheric absorption measurement [Davis 1990, Rawlings 1976]. This is where the strength of a continuous wave (CW) signal from a remote transmitting station is monitored in time using a calibrated signal strength meter. The strength of the signal was expected to increase for most frequencies and propagation paths as the absorbing D layer of the ionosphere disappeared with the loss of sunlight (as shown in Table 1).

### *A DESCRIPTION OF THE RADIO AMATEUR AND RA EXPERIMENTS*

In detail the experiment required noting down the strength of radio signals to and from the Continent, before, during and after the eclipse. That meant between about 8:30am and 13:00pm (British Summer Time) on the 11th August 1999 and similarly on a control day either the day before or after. Some used computers to record the signal levels but many just used pen and paper. One problem, that was identified early on, was that most signal strength meters or dials on radio amateur equipment are likely to be too unreliable to be used directly for measuring received signal strength. However advice was provided in detail as to how to overcome this by supplying a procedure as to how to calibrate these meters into dB. This many, but not all, took the time to do. However even uncalibrated measurements had value. The importance to take great care about the exact timing of the observations was constantly emphasised.

Publicising the experiment to radio amateurs was done at Conferences, through the HF Technical Working Party, speaking at local clubs and publishing articles in radio amateur journals such as RadCom and Radio Today. Laminated certificates were sent to the radio amateurs who sent their results to Rutherford Appleton Laboratory as thanks.

## RESULTS FROM THE RADIO AMATEURS

Figure 13 shows just a selection of the responses sent in by the radio amateurs. The frequencies chosen here range from 864kHz to 7MHz. The change in signal strength for these frequencies ranges from 10 to 40 dB. Similar to the ITU day-time/night-time values shown in Table 1. In Figure 14, Figure 15 and Figure 16 just three examples of the variations of signal strength in time are shown along side a map of Europe with the locations of the transmitters and receivers in each case. The path the eclipse totality passed is indicated on the map by the thick black line. In the first example (Figure 14) the signal strength variation is for a propagation path which must traverse the path of totality. The propagation paths are shown as simple straight-line projections onto the ground on the map, though the radio signals must have reflected from the ionosphere and in reality follow a 3D path that would require a ray trace to model accurately. The frequency was 6.065 MHz and was recorded in Spain from a transmitter in Sweden. The 100% eclipse shadow passed the mid-point of the propagation path at approximately the same time as the signal strength reached its maximum. A red star indicates this point on both the map and time plot. The slight difference in time is probably due to a combination of the sampling interval and the fact that the eclipse path on the map is shown for the eclipse on the ground. Up at ionospheric altitudes the eclipse occurred a few minutes earlier.

The maximum signal when the eclipse is at the mid-point is the behaviour one would expect. That is although the point of reflection of the radio waves is going to be the E layer above the D layer, the maximum increase in signal strength of a received signal would occur when the average absorption is a minimum on both upward and downward legs of the journey. This is illustrated in Figure 12.

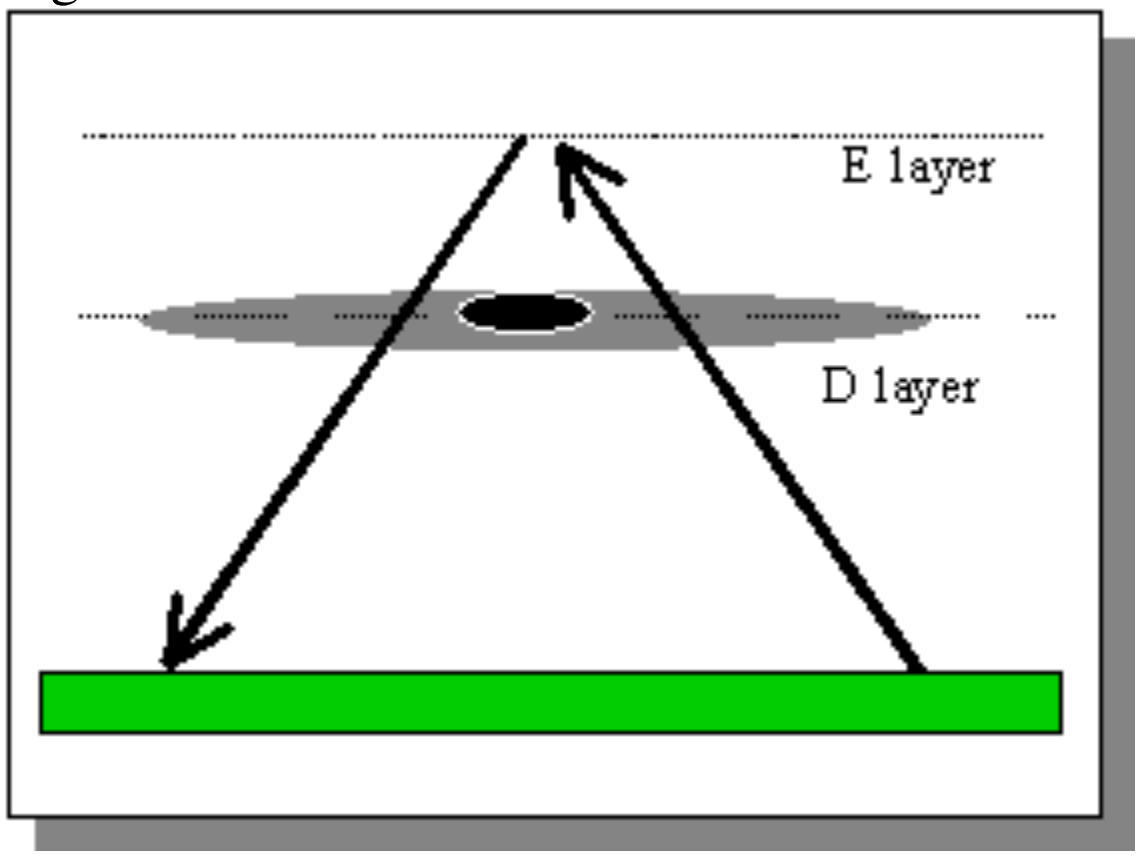

*Figure 12. A sketch showing a simple view of radio waves reflection off the E layer of the ionosphere. The eclipse 100% shadow is shown as the dark oval in the centre as if it were effecting the D layer. The shaded oral represents the partial eclipse shadow that is effecting the D layer absorption on the upward and downward propagation.*

A similar observation was found for most frequencies and propagation paths across and along the path of totality. However there were exceptions which are discussed in a later section . Two examples for the variation in signal strength at 3.522MHz and 864 kHz from a receive site directly under the path of totality in the UK from transmitter just north of the totality are shown in Figure 15 and Figure 16. Here too the maximum signal occurs approximately when the eclipse totality was at the mid-point of the projected straight-line propagation path.

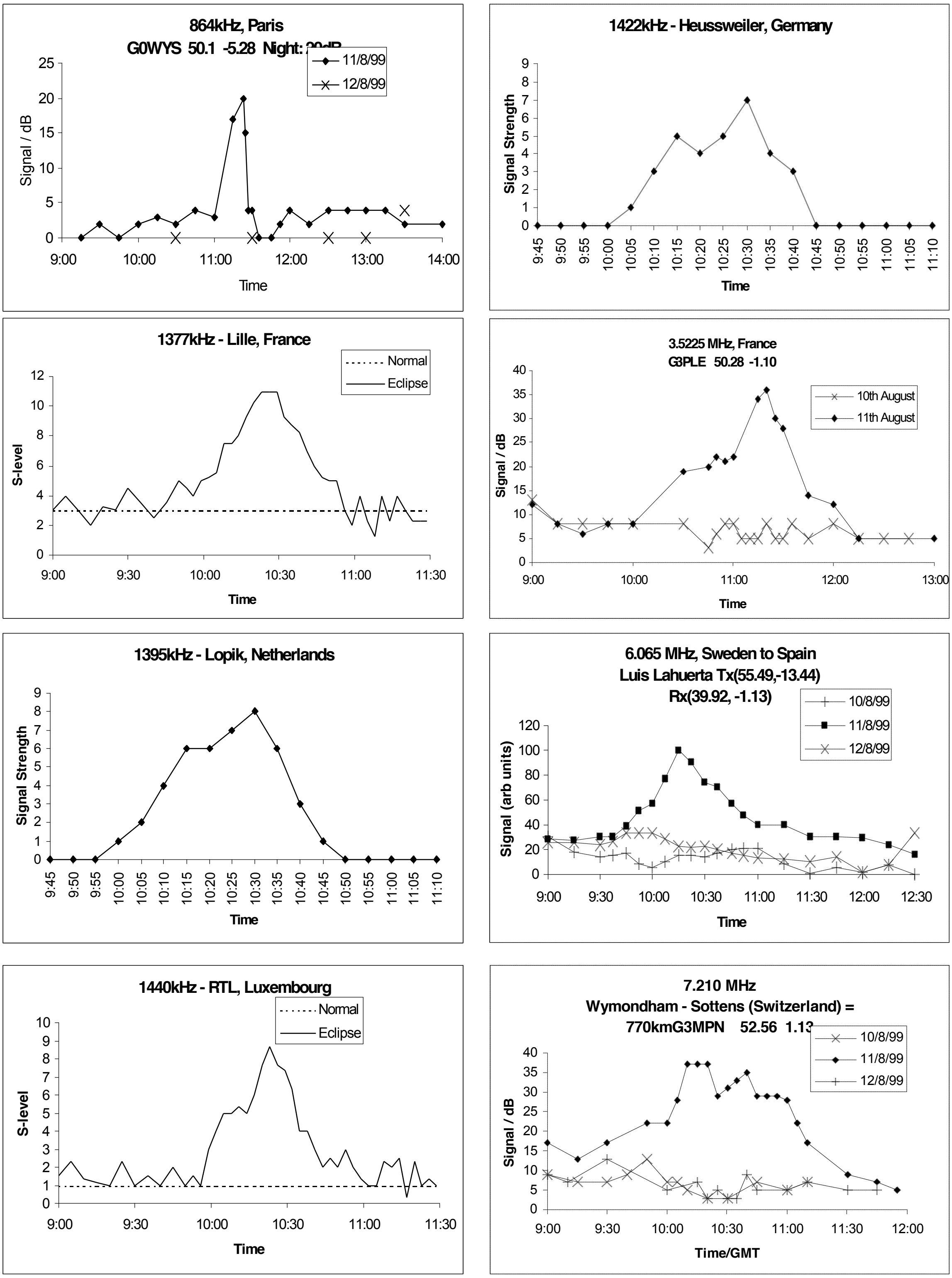

*Figure 13. A selection of the observations during the 1999 solar eclipse of the signal strength of CW transmissions from Continental stations received in the UK as recorded by members of the Radio Society of Great Britain (RSGB).*

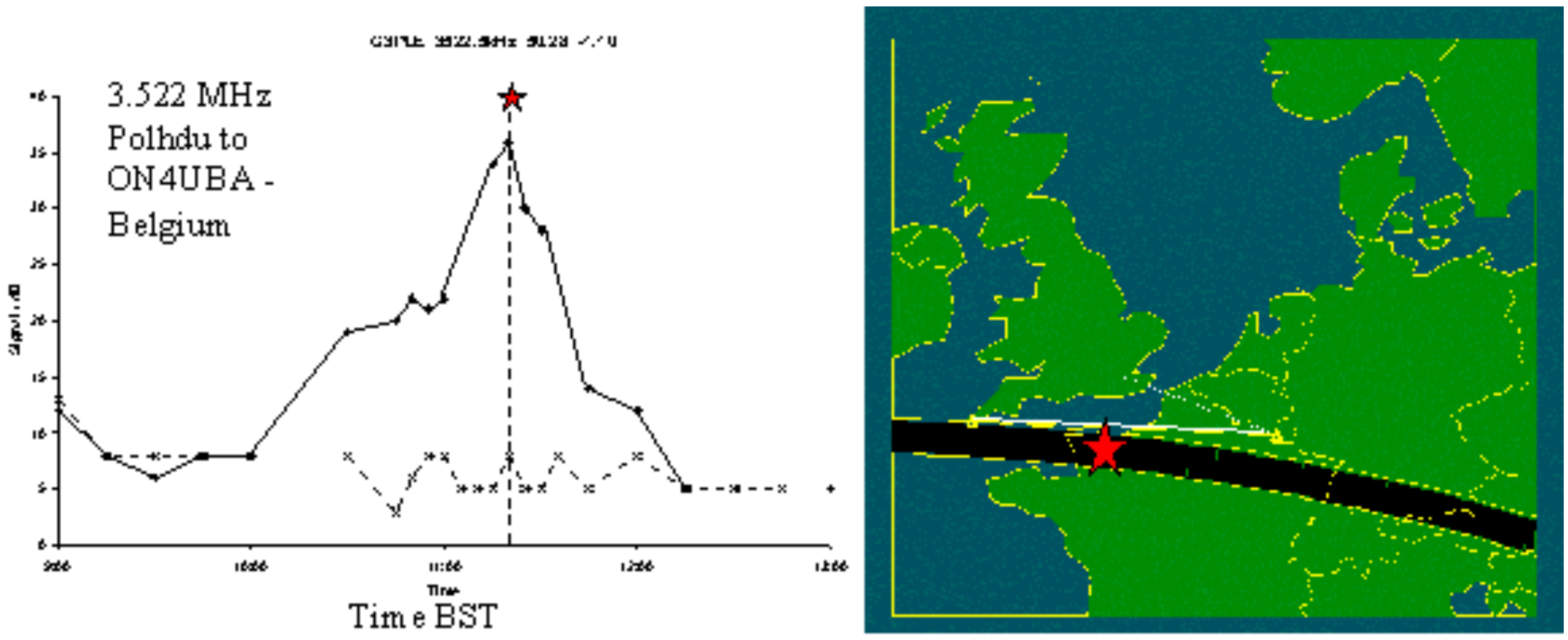

*Figure 14. The variation in signal strength for propagation across the path of totality at 6.065 MHz.*

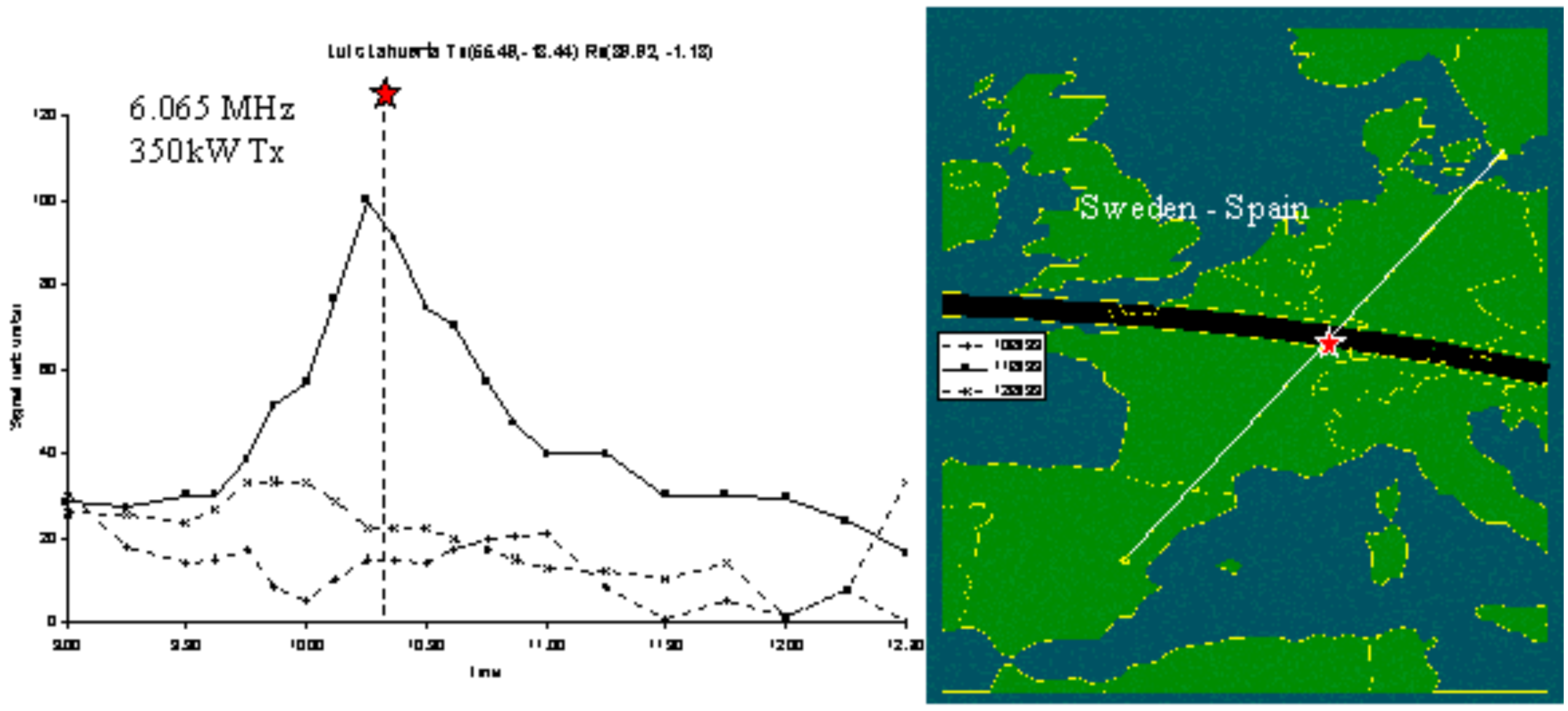

*Figure 15. The variation in signal strength for propagation along the path of totality at 3.522MHz.*

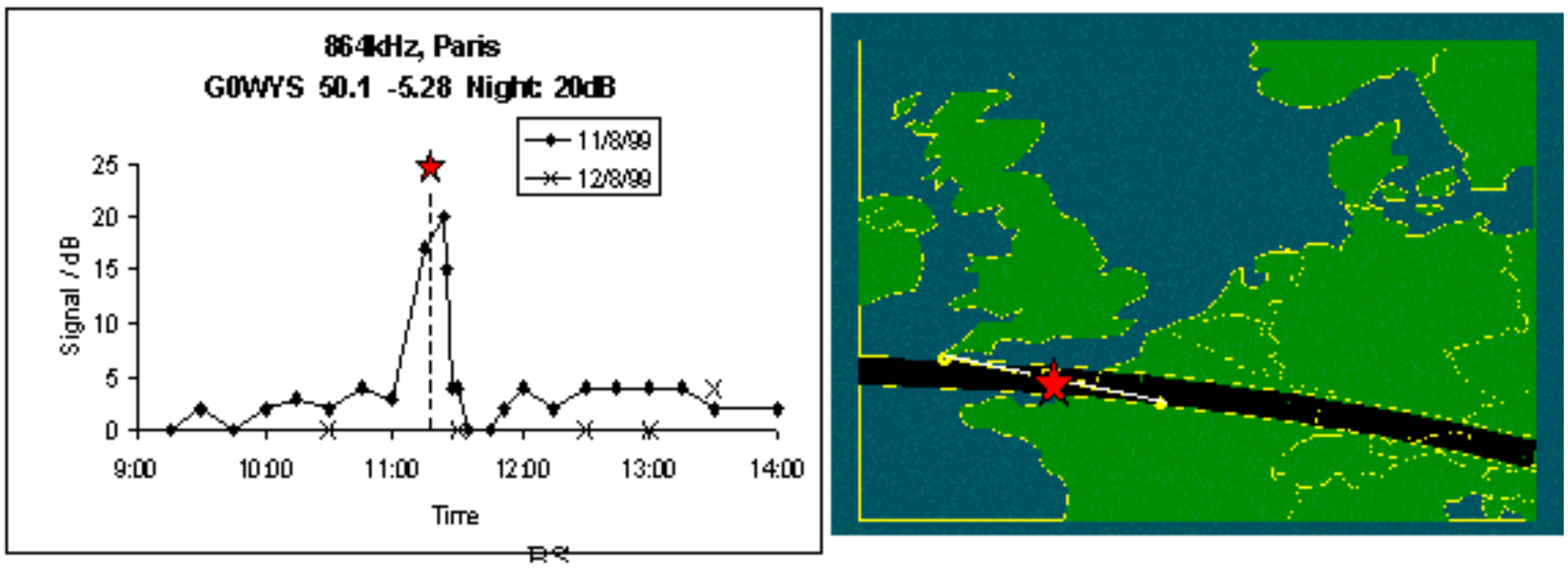

*Figure 16. The variation in signal strength for propagation along the path at 864kHz.*

## *LF AND VLF OBSERVATIONS DURING THE ECLIPSE*

The LF and VLF band (3kHz -300kHz) data from radio amateurs showed some very interesting effects of the eclipse. The plots in Figure 17, which show the temporal variation in the signal strength at 75kHz from the Swiss time clock transmissions (HBG), illustrate this. These are very different from the responses at the MF and HF (300kHz – 30MHz). The map in Figure 18 shows the different direct line propagation paths between the transmitter and the receiving locations for the signals shown in Figure 17. There is more than one explanation for the oscillations clearly seen at this frequency. Firstly they could be the result of phase changes incurred by the radio signals due to the effects of the eclipse shadow on the ionosphere where the signals are being reflected (this has been seen before). Alternatively changes in the heights of the layers and multi-path interference are two other explanations or a combination of all of these effects. At these frequencies the radio waves are predominately being reflected by the D layer of the ionosphere which is undergoing a lot of changes as a consequence of the eclipse.

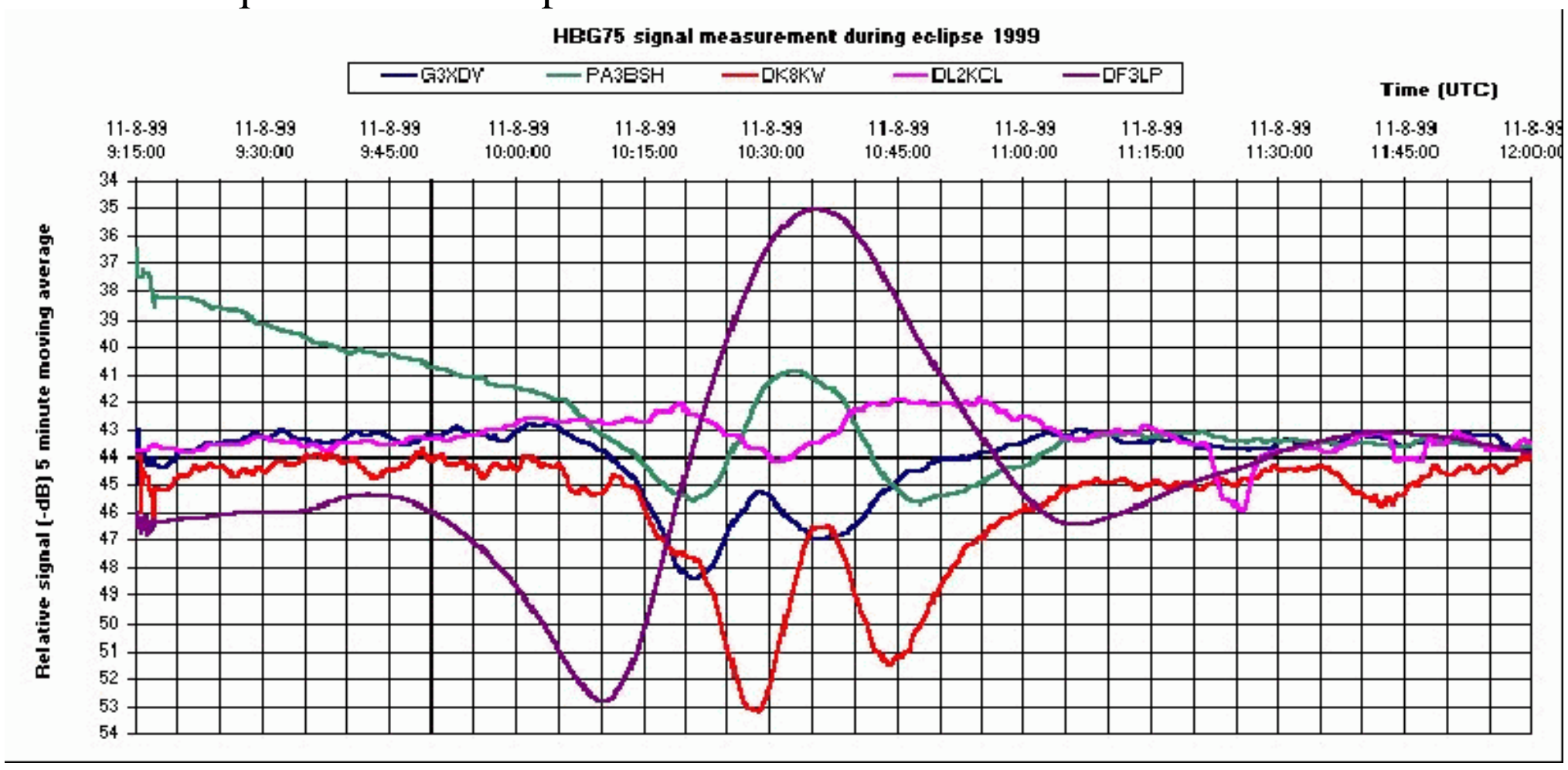

*Figure 17. A plot of the variation in signal strength at 75 kHz (HBG time clock) observed from a variety of locations across Europe. [M. Sanders, 1999]*

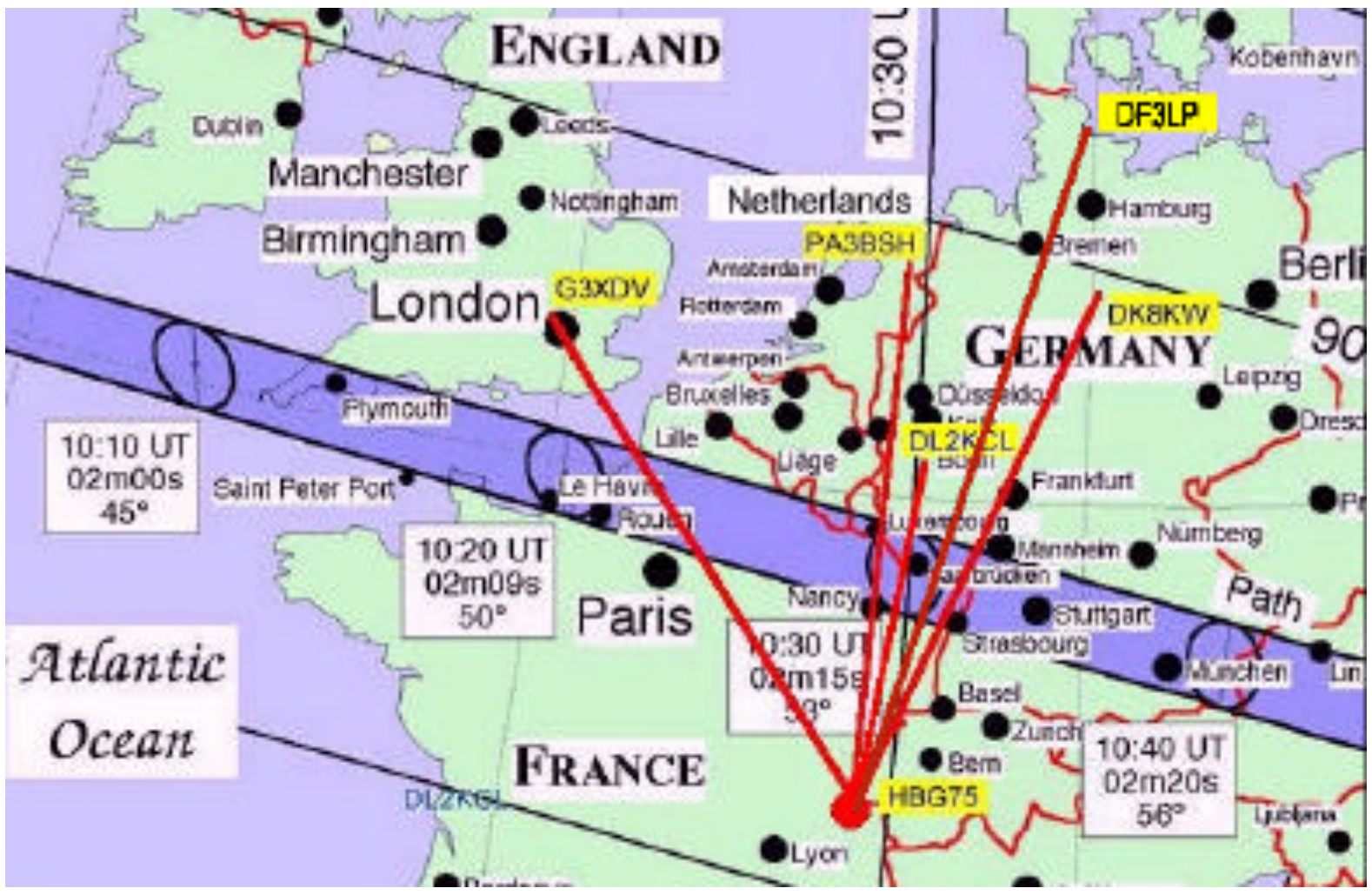

*Figure 18. The location of the 75kHz HBG Swiss time signal transmitter and the locations of the receiving radio amateurs identified by their radio designations.*

### *CONTRIBUTIONS FROM THE RA REGIONAL OFFICES*

Staff at the RA Regional Offices and at the Baldock monitoring station also contributed to the radio propagation observations of the total solar eclipse in 1999. A total of 16 computer-controlled radio “scanners” were purchased and distributed to the RA regional offices to make accurately timed observations using computers. The receivers used were ICOM/IC-PCR1000 programmable LF/MF/HF radio receivers controlled by PC with a 10-second sampling interval. Each receiver was calibrated into dBm before deployment at a range of spot frequencies from 250kHz to 15MHz. The antennas used were mainly low noise, broadband active whip antennas (SONY AN-1). These provided a more consistent data set of observations.

# PROPAGATION AT 1440KHZ

One of the best transmitter stations to observe turned out to be the 1200 kW transmitter of Radio Luxembourg broadcasting at 1440kHz from Marnach. An example of the signal strength variation on the eclipse day and a control day is shown in Figure 19. The decrease in signal strength due to the increasing levels of absorption in the D layer as the sun rises at dawn is well illustrated on both days between 4am and 6:30am. During the 1.5 hours of the eclipse the signal level can clearly seen to return to 60% of the night-time (top panel). The straight-line projection onto the ground of the propagation path from Marnach where Radio Luxembourg to Birmingham is show as the dashed line in the map in Figure 20 along with paths to 5 other RA regional offices (solid lines) and Chilton, Rutherford Appleton Laboratory (dotted line).

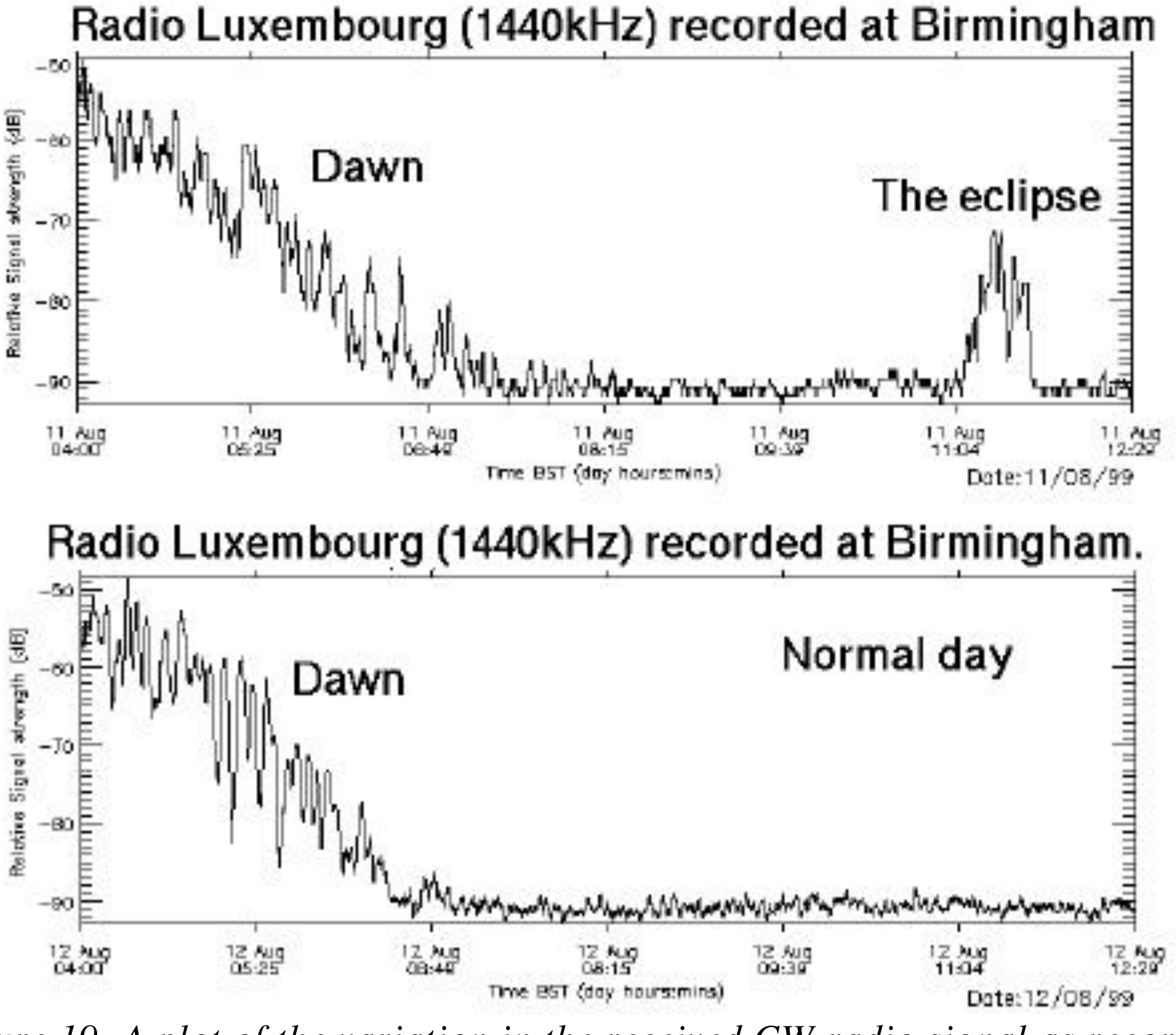

*Figure 19. A plot of the variation in the received CW radio signal as recorded in Birmingham RA Regional Office in the UK of the 1440kHz (±1.4kHz) carrier emanating from Radio Luxembourg at Marnach (a) for the morning of the total solar eclipse and (b) the day after the eclipse.*

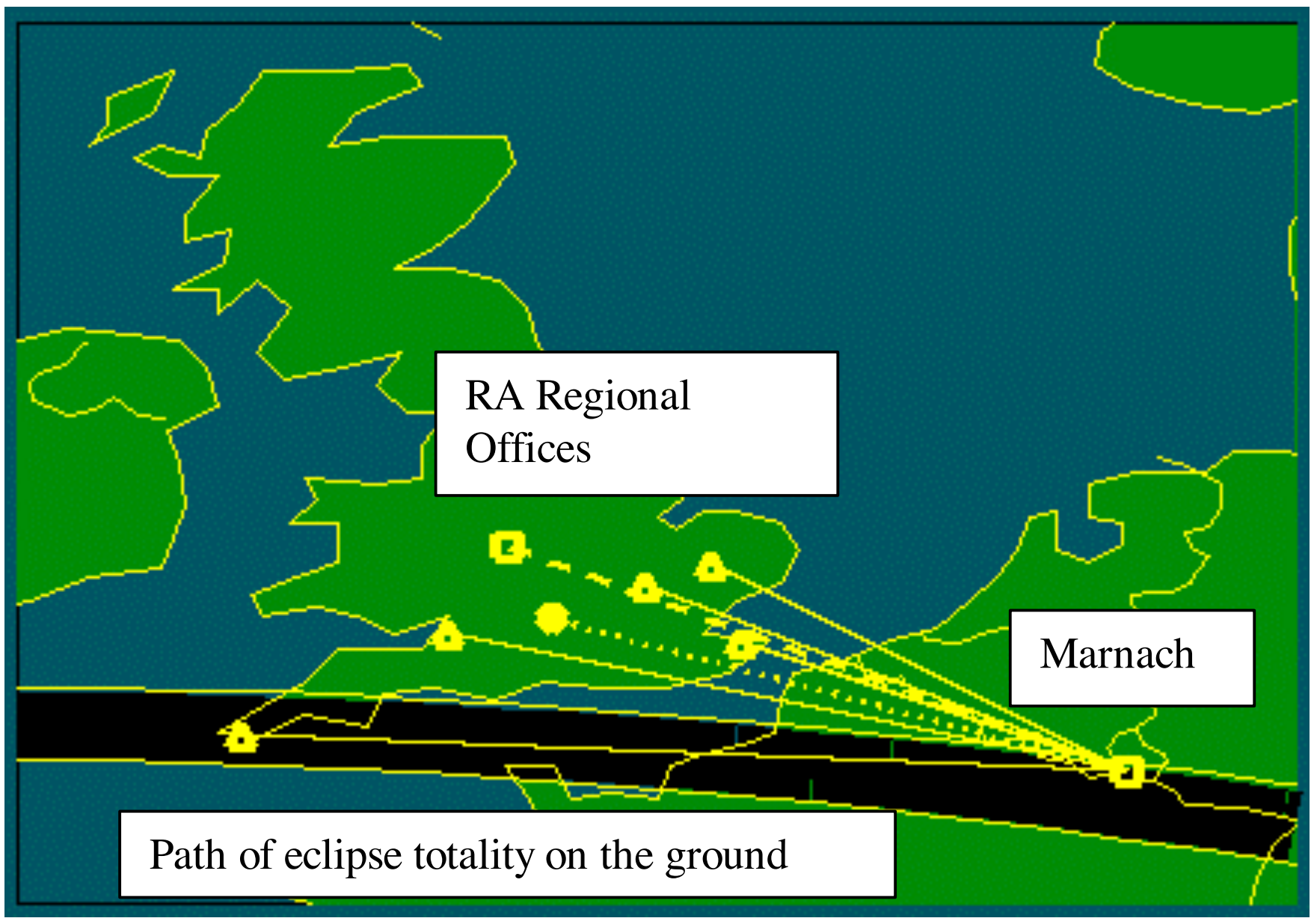

*Figure 20. A map of Europe with the path of totality (at ground level) of the solar eclipse and the location of the 1440kHz Marnach transmitter in Luxembourg at the northern edge of the eclipse path of totality and the receiving stations in the U.K.*

Figure 21, shows the temporal variation the signal strength, similar to the plot shown in Figure 19, but with the signal level for the normal day subtracted to highlight the difference made by the eclipse. This type of plot allows a closer examination of the timing of the effect on the radio reception and the passage of the lunar shadow to be compared for all the receiving stations. The times and the peak values for the all the receiving stations are listed in Table 1. The latitude and longitude of the receiving stations and the great circle ground range between the transmitter and each receiver is also included in the table. The precision of the receiving station clocks was checked manually before and after the time of the eclipse and corrected for.

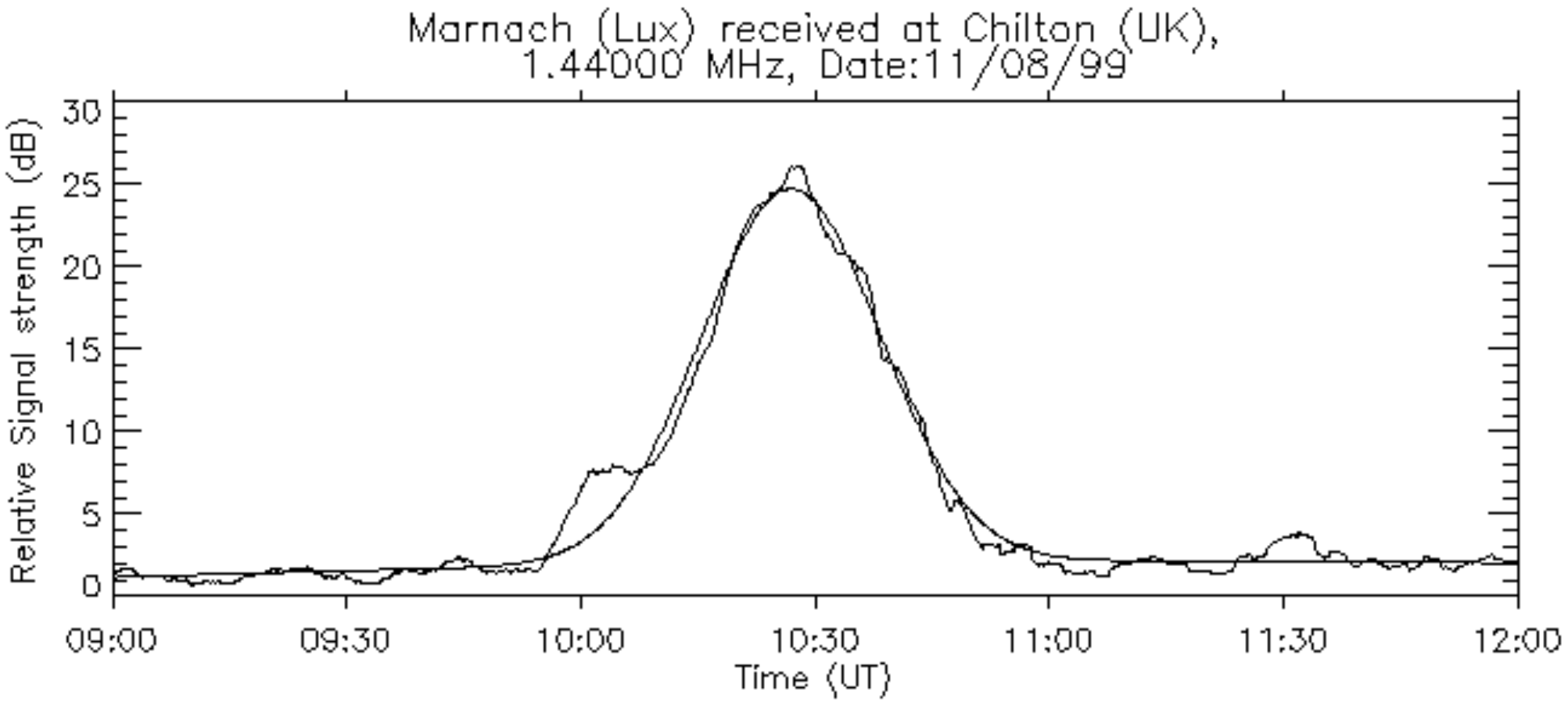

*Figure 21. Variation in received signal at Chilton in the UK of the 1440kHz CW Radio Luxembourg carrier broadcast from Marnach (Radio Luxembourg). Here a 5 minute smoothing has been applied.*

*Table 2. The results from the UK receivers monitoring the carrier frequency from the Marnach transmitter. The 100 % passed over the Marnach transmitter at 10:28:58 UT [Bell 1990]. * Uncalibrated receiver. Marnach Tx is* 49.62N, 6.0 E.

| Receiver Station Rx | Lat. | Long | Distance (km) Tx- Rx | Time of max eclipse Rx | % max eclipse Rx | Time of signal maximum (UT) | Rise (dBm) |
|---|---|---|---|---|---|---|---|
| Baldock | 52.00 N | -0.13 E | 505.14 | 10:19:37 ¥ | 95.3 | 10:28:24 | 23.8 |
| Birmingham | 52.49 N | -1.89 E | 636.37 | 10:17:57 | 93.5 | 10:26:44 | 25 |
| Bristol * | 51.4 N | -2.64 E | 641.50 | 10:16:25 | 97.3 | 10:28:07 | 39.9* |
| Bury St Edmunds | 52.25 N | 0.72 E | 471.30 | 10:21:54 § | 94.3 | 10:28:30 | 20.3 |
| Canterbury | 51.28 N | 1.07 E | 394.68 | 10:21:31 | 97.0 | 10:26:08 | 21 |
| Chilton | 51.56 N | -1.3 E | 558.17 | 10:18:24♦ | 95.9 | 10:27:53 | 23.9 |
| Helston | 50.1 N | -5.27 E | 808.31 | 10:12:00 | 100.0 | 10:24:33 | 28 |

The eclipse totality was directly over Luxembourg at 10:28:58 UT. This is very similar to the time of maximum signal (column 7 in Table 2) rather than the time of local eclipse maximum over the receiver stations (column 5 in Table 2) for all but one case. What this table reveals is the general tendency for the peak signal strength at 1440kHz to occur when the eclipse shadow was much closer to the transmitter than either the path mid-point. This was not the case for HF frequencies. The exception to this is the case for 1440 kHz being received at Helston where the receiver was also directly under the path of totality and the propagation from the transmitter to the receiver would have experienced the maximum effect of the eclipse.

What was most unexpected was that for the signal strength responses are centred on the time the eclipse passed over the transmitter, the enhancement of > 10dBm to the reception in the UK continues when the lunar shadow had progressed well into southern Germany, more than 10 minutes after the maximum.

This suggests that the loss of absorption directly over the transmitter had a more significant effect than the loss of absorption at any other region of the path for this transmitter for these cases at 1440kHz.

# OBSERVATIONS MADE BY THE IONOSONDES ON THE IONOSPHERE

The monitoring ionospheric radars or ionosondes that are funded jointly by Particle Physics and Astronomy Research Council (PPARC) and RA made direct observations of the effect on the ionosphere of the eclipse. These instruments transmit HF radio waves vertical upward at the ionosphere from 1MHz to 30 MHz and record the time of flight of the return echoes [Davis 1990]. From this the maximum or critical frequencies of each of the ionospheric layers can be measured.

The observations of the change in the critical frequencies of the F and E layers from an ionosonde are shown in Figure 22. This ionosonde had been moved to Cornwall so as to be directly under the path of totality. The vertical line indicates the time of local eclipse maximum.

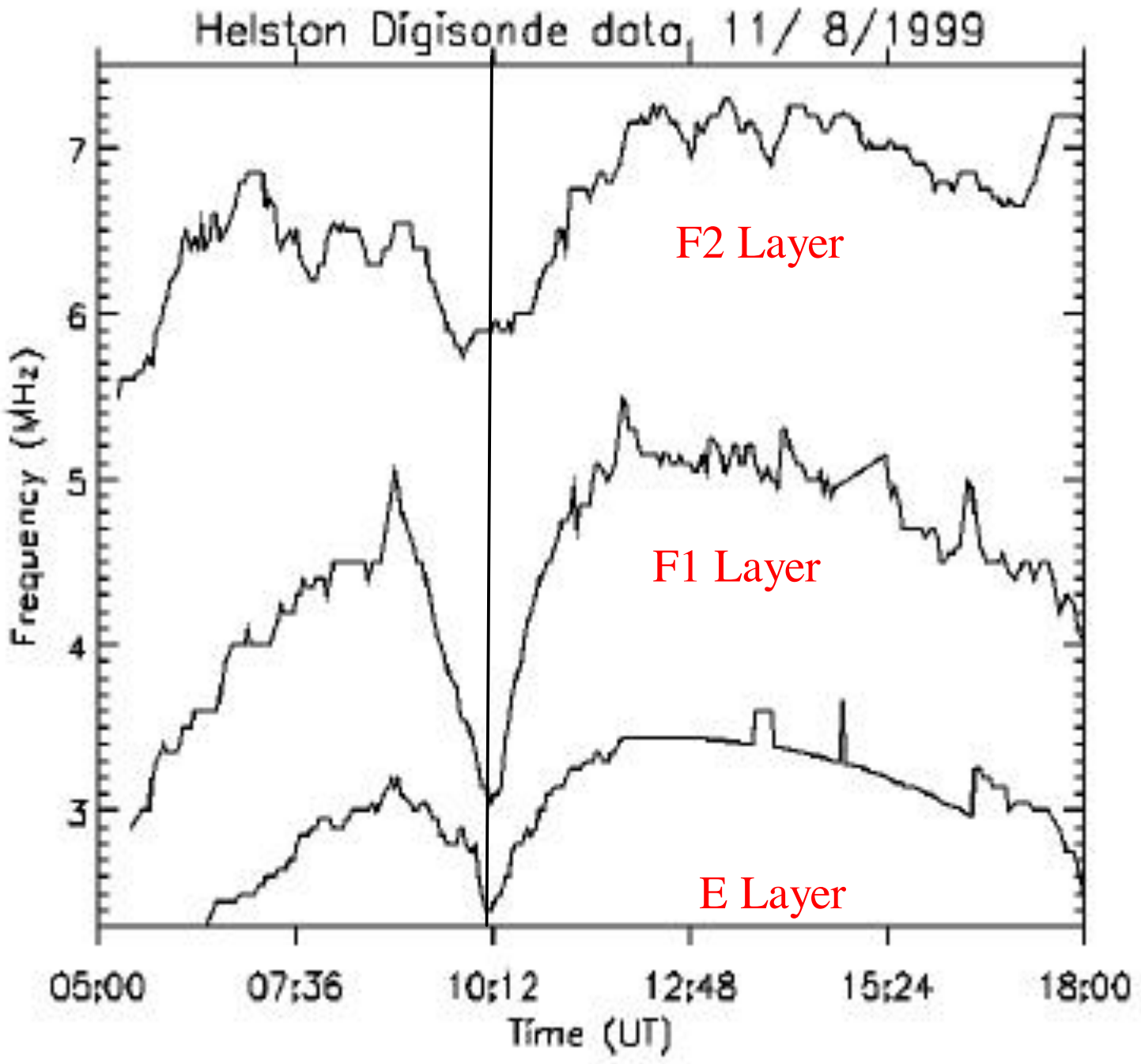

*Figure 22. Reduction in each of the ionospheric layers during the eclipse as observed directly under the path of totality at Helston in Cornwall.*

All the critical frequencies drop significantly during the time of the eclipse, which is in progress from 8:56 UT to 11:31UT, with the maximum at 100% totality occurring at 10:12 UT (on the ground). The level of the drop in electron density for the layers is between 20 to 35% of the normal daytime values. From the plot it can be seen that the sharpness of the reduction in the ionospheric layers was more significant for the E layer than for the higher altitude F layer. This is because of the stronger solar dependence of the E layer compared to the F layer and because the eclipse was total for the E layer but not total for the F layer. This comment is explained by referring to the map in Figure 23. Allowance has to be made for the declination of the sun and the altitude of the ionospheric layers. In Figure 23 the latitude and longitude of the path of totality on the ground (black) and at the altitude of the D, E and F layers of the ionosphere (red, orange and yellow) are projected onto the map. In any analysis of the ionosphere and radio propagation this geographical projection of the lunar shadow onto the ionosphere had to be accommodated.

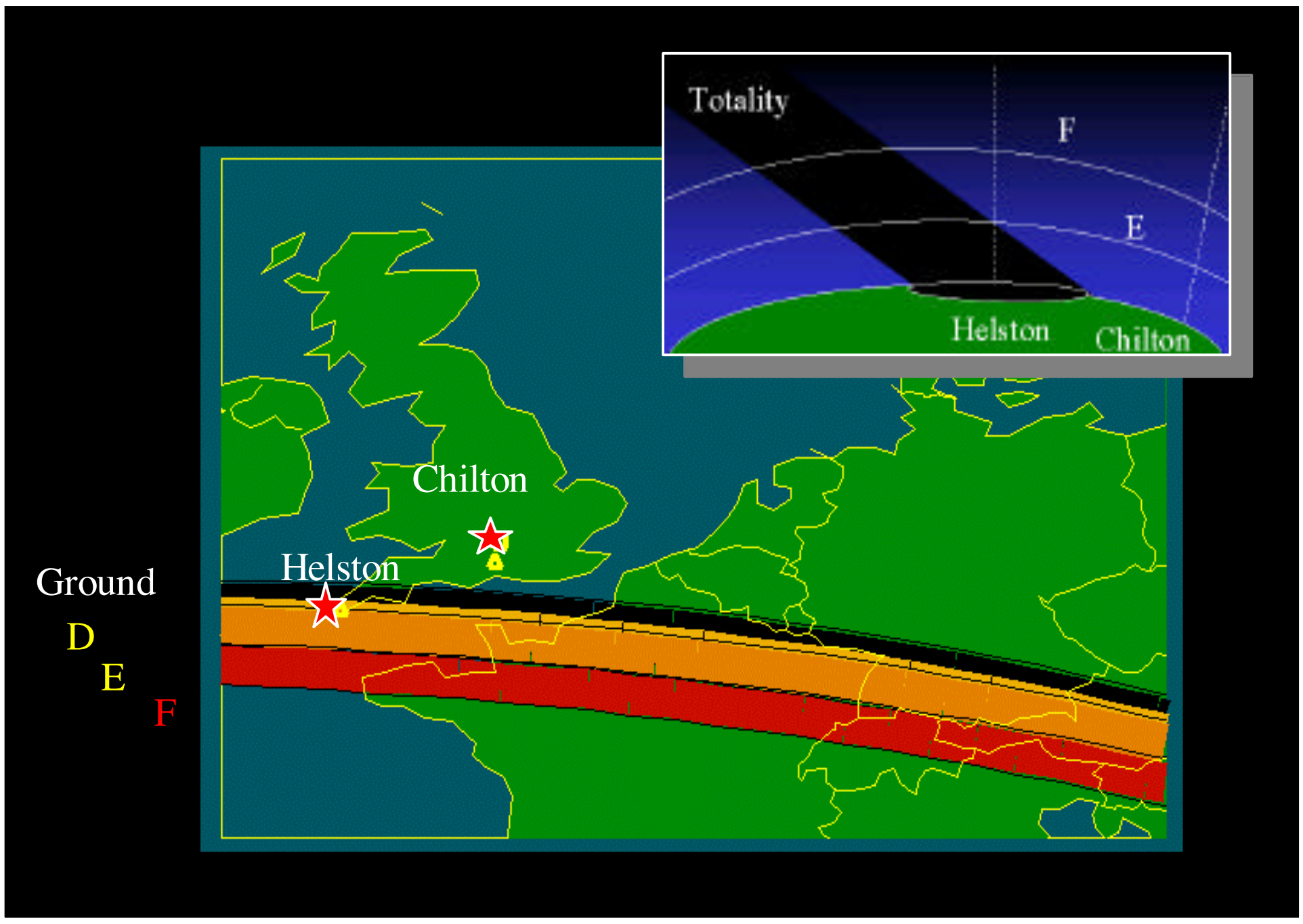

*Figure 23. A map showing the path of totality projected onto the altitude of the ionospheric D, E and F layers at 50, 100 and 250km. Also show is the plan view showing the relative location of the ionosonde stations to the inclined lunar shadow.*

# ANALYSIS AND MODELLING OF THE RESULTS

Although some initial analysis has already been shown in the previous sections along with the results. The following sections aim to examine some of the general observations and determine the physical origins.

## REDUCTION IN THE MUF AND THE EFFECT ON PROPAGATION

Not all the observations of the signal strength showed an increase. The two examples below (Figure 24) clearly showing a reduction of the order of probably 30dB or more (data here is uncalibrated but 1 "S-level" generally is about 5dB). This however is explained by changes in the critical frequency of the reflecting layer, which is the F layer of the ionosphere in this case.

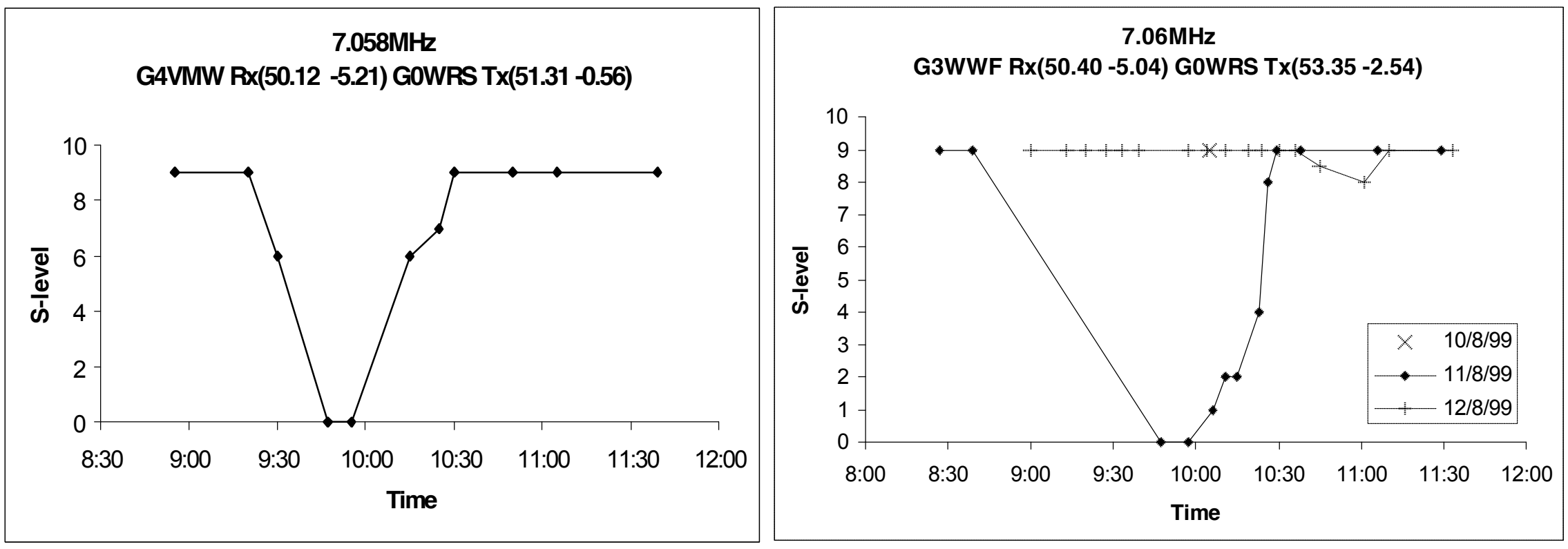

*Figure 24. Two example where the variation of signal shows a reduction rather than an increase during the eclipse.*

When radio signals propagate close to the maximum frequency that the specific ionospheric layer will allow reflection, then other factors than the D layer absorption effect the signal strengths. This is the maximum usage frequency (MUF) for a specific point to point communication link. This different type of absorption is the so called deviative absorption, caused by the radio waves *refracting* close to a *resonant frequency* of the ionosphere. The map in Figure 25 shows how for these two signal strength examples the propagation paths were short (about 200km). The location marked by the star on the map shows the location of the Chilton monitoring ionosonde radar. This instrument makes direct measurements of the changes in the critical frequencies of the ionospheric layers that can be translated into MUFs for practical communications use. The geographical proximity of the mid-point of propagation direction of the signal strength, which is the probably point of reflection, and the ionospheric measurements at Chilton are very close in this case. This allowed the changes in the MUF to be calculated with accuracy using the ionosonde data for the specific paths in question for the period of the eclipse. This is shown in Figure 26. Here the MUF prior to the eclipse was above 7.5 MHz, which is above the 7MHz working frequency being used for the signal strength measurements Figure 24. However the MUF drops below 6.5 MHz as a consequence of the eclipse shadow passing resulting in the drop and temporary loss of signal.

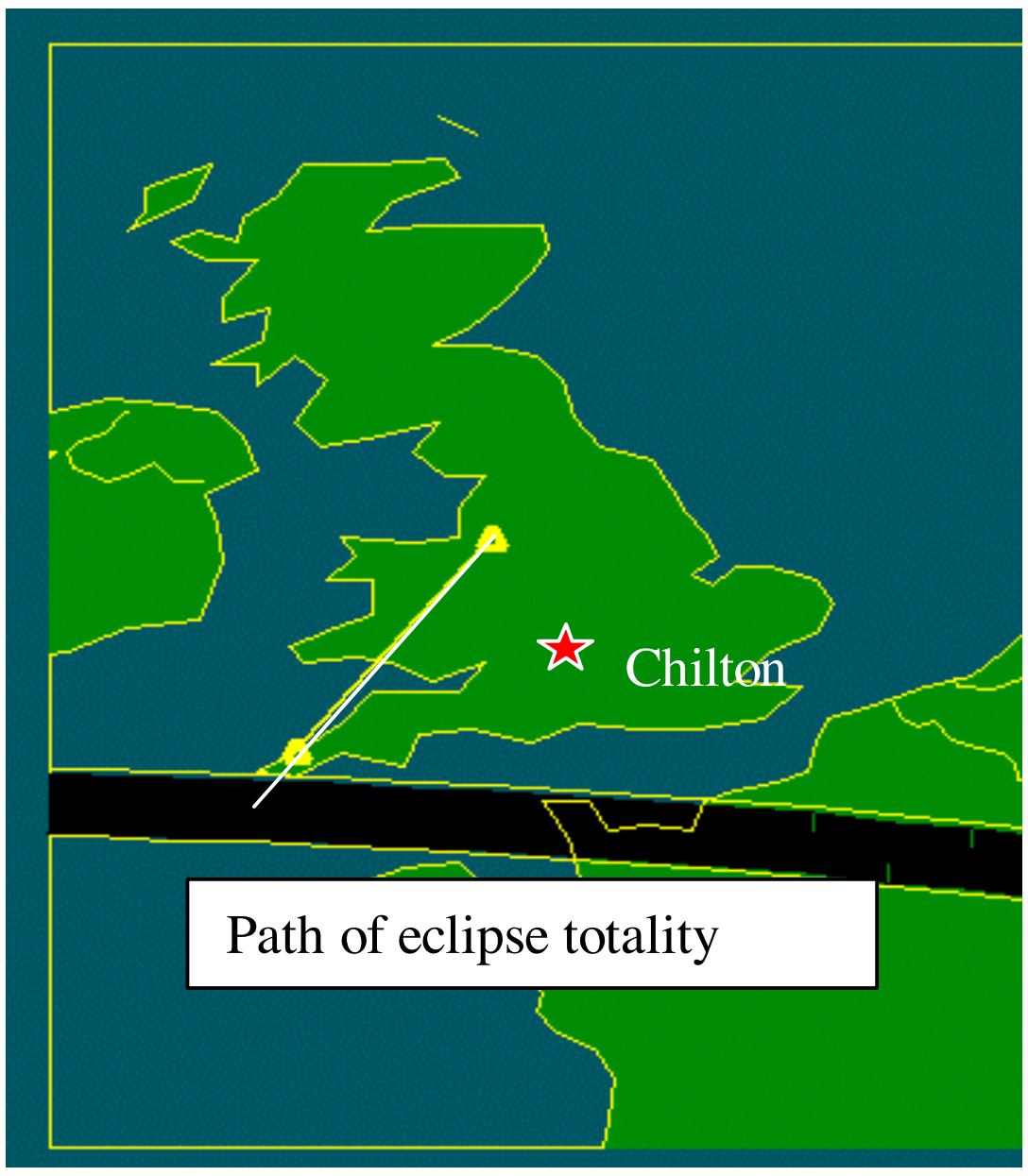

*Figure 25. A map showing the location of the receiving and transmitting stations and the Chilton ionosonde relative to the path of totality.*

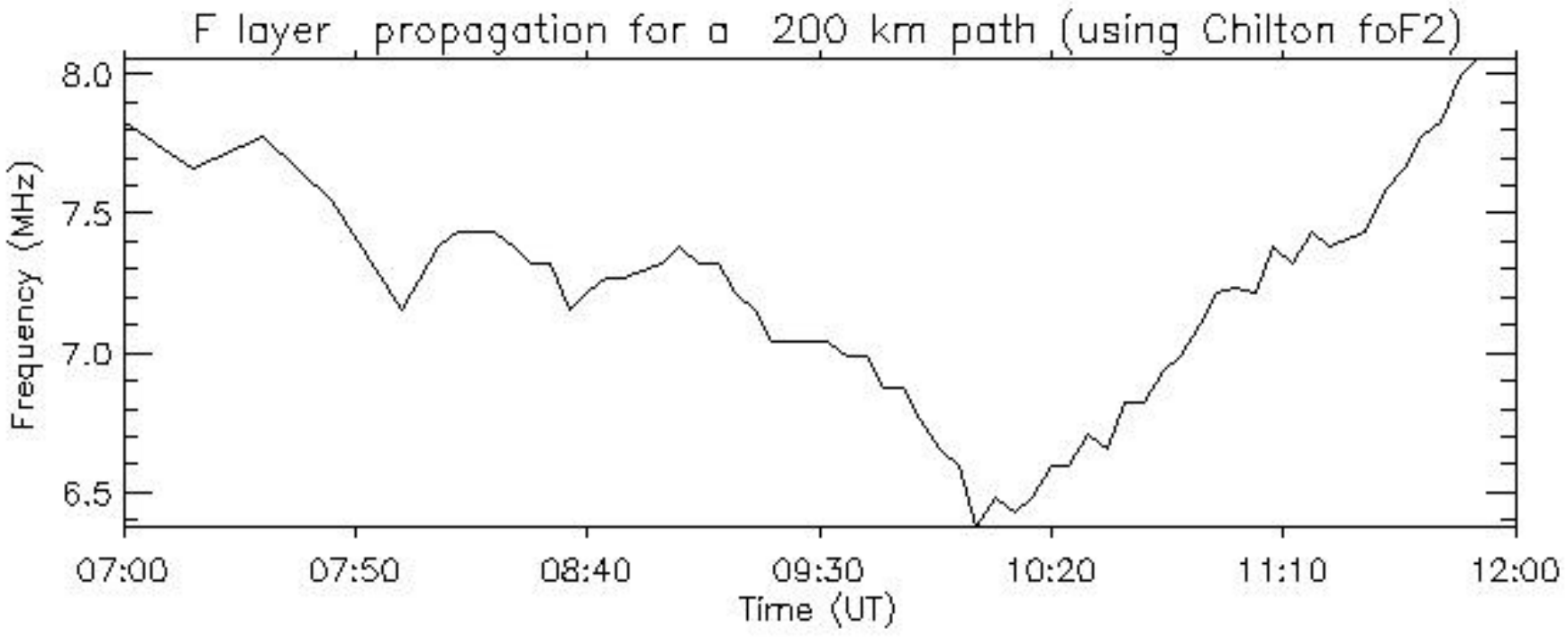

*Figure 26. The calculated changes in the maximum usage frequency (MUF) for the propagation paths shown in the previous images based on Chilton ionosonde data.*

# TEMPORAL VARIATION OF THE ECLIPSE EFFECT AND SOLAR EMISSION

In examining the temporal variations in the signal strength, which is effectively a measure of the ionospheric absorption, shown in Figure 13 for many frequencies and paths, some common elements are apparent. Most significantly is that on both the rising and the falling edge there is generally a slight hesitation or change in the rate of change of the loss of absorption. Another example of this characteristic is shown in Figure 27. A similar, but inverted profile is seen on the lowermost E layer trace in Figure 22.

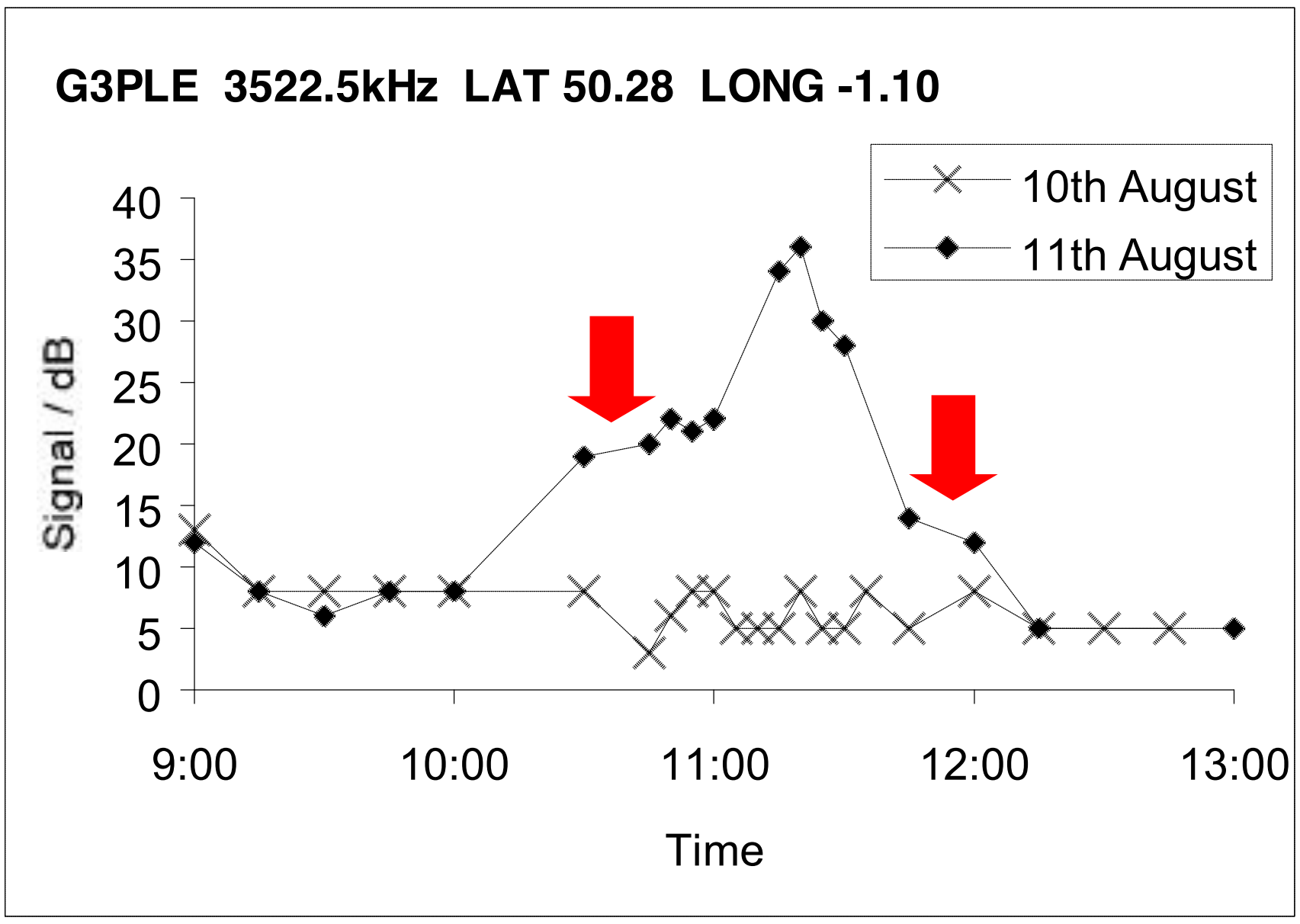

*Figure 27. An example of the variation of signal strength change observed on many of the radio eclipse propagation measurements with a change in gradient both before and after the eclipse maximum.*

This could have two causes. The first is due to the selective obstruction of bright regions or prominences on the sun's disk which are emitting higher levels of extreme ultra violet and /or x-rays. The second reason is that it is due to changes in the ionosphere.

By using data from the SOHO (Solar Heliospheric Observatory) satellite these two effects can be separated. An image of the sun taken in EUV (17.1nm) for the morning of the eclipse is shown in Figure 28. This is the wavelength most closely available to the wavelengths that effect the ionosphere. In the past ionospheric studies during eclipses generally assumed a uniformly illuminated solar disk. These images clearly show the non-uniformity of the sun's emission across the solar disk.

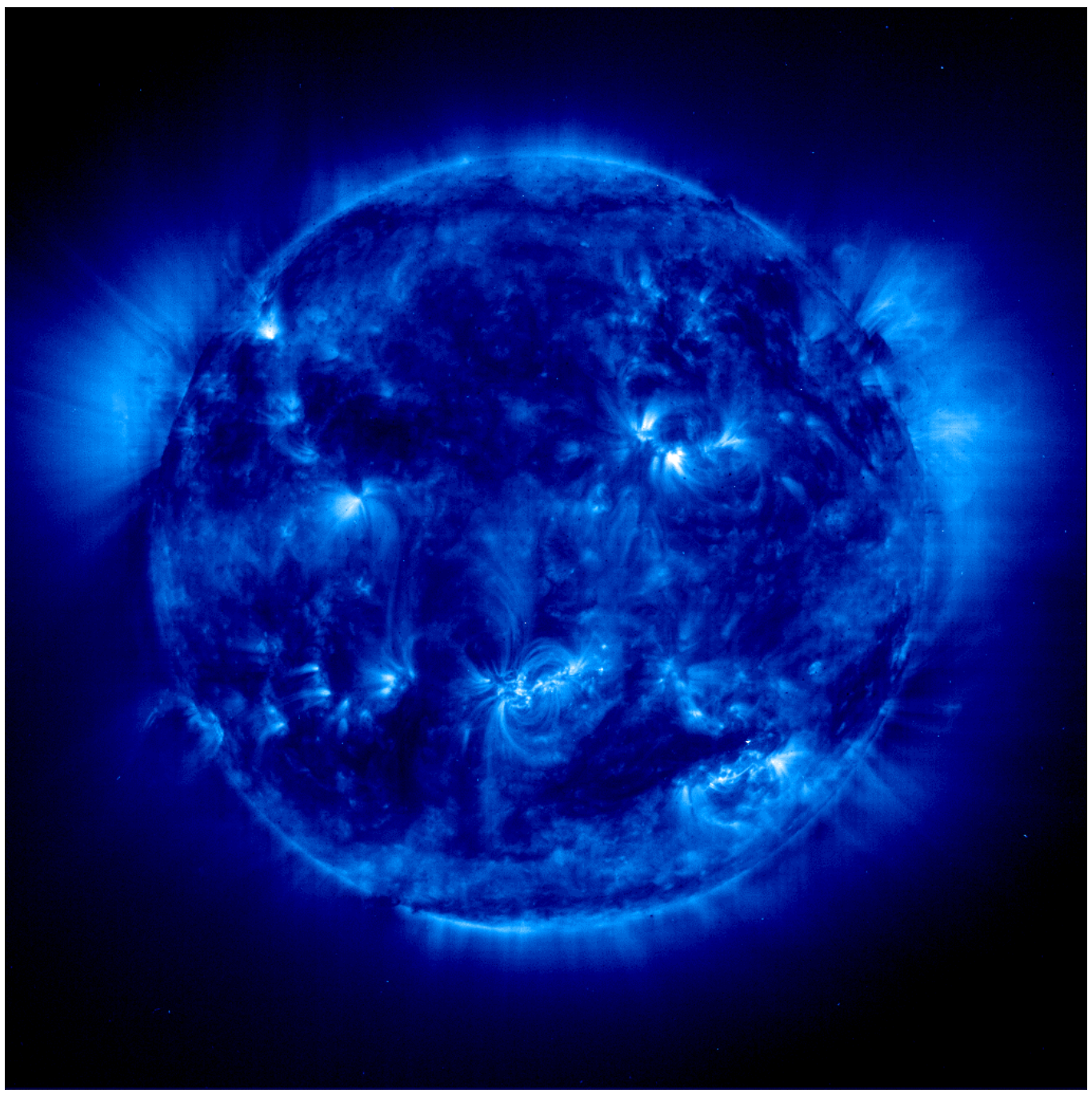

*Figure 28. A SOHO image of the sun taken in extreme ultraviolet wavelengths (17.1nm) at 10:12 UT on August 11th 1999. (Thanks to SOHO EIT).*

The occasion of an eclipse the moon is progressively obscuring the solar disk s seen from the ground. One can then simulate this very easily, using SOHO images such as the one shown in Figure 28, using the exact details of the relative size of the sun and the orientation of the SOHO image and the eclipse local circumstances [Bell 1996].

This provides a temporal plot of the obscuration of the solar prominances that is plotted below (Figure 29).

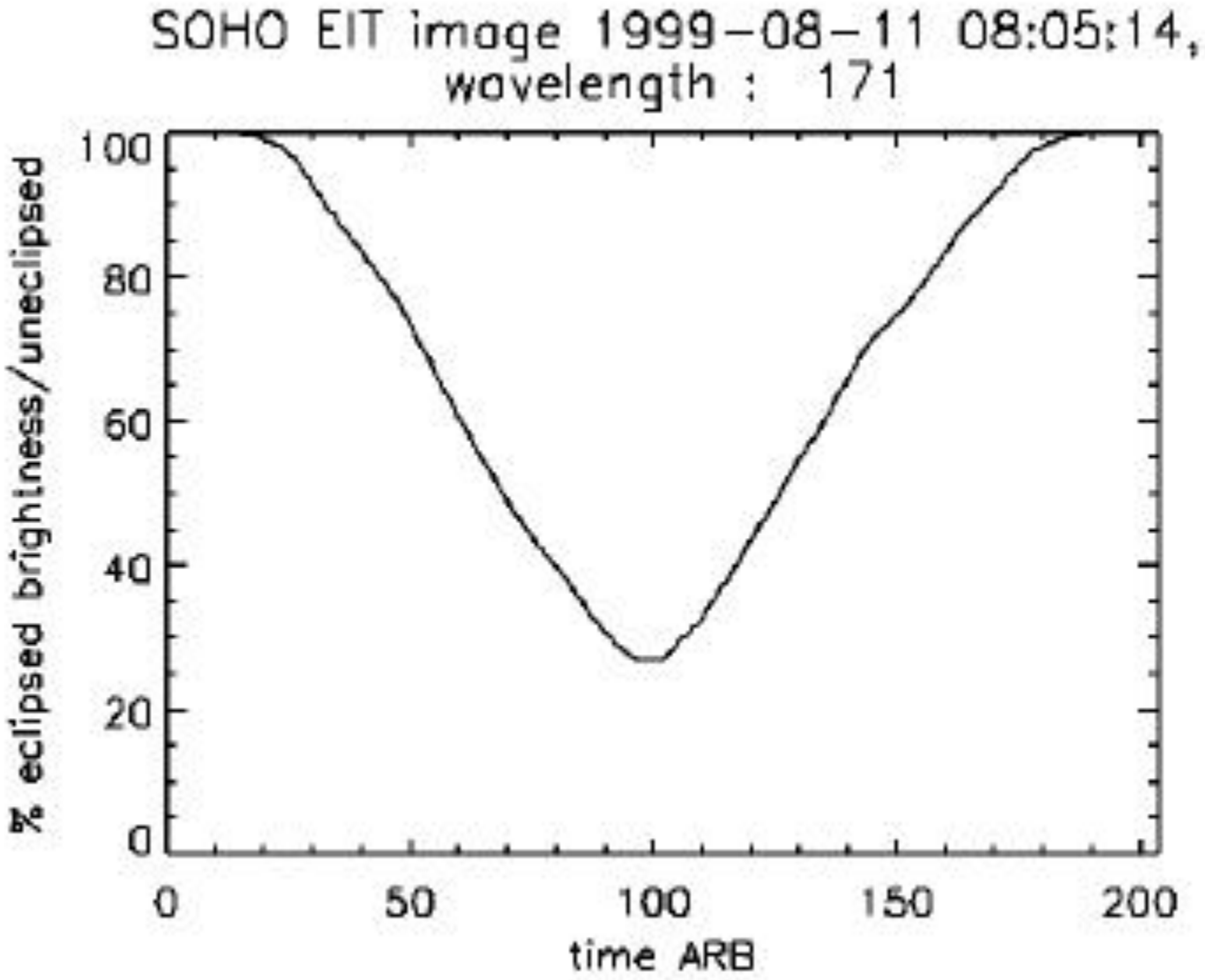

*Figure 29. The change in the fractional EUV illumination derived from EIT SOHO data at 17.1nm relative to an unobscured solar disk.*

What is surprising is that despite non-uniformity of the emission visible in the image in Figure 28, the variation in Figure 29 is so smooth. Allowance was made for the dynamic response of the detectors and the distortions of the imaging process. What this analysis suggests that the variations observed on the radio propagation and ionosonde observations have to be predominately ionospheric in origin rather than solar. The observations of the ionosondes shown in Figure 22 show that the ionosphere did change in a similar way to the propagation responses. To understand this response is the realm of ionospheric modelling.

# MODELLING THE IONOSPHERIC RESPONSE

The attempts to model the ionospheric response to the disturbance of the eclipse is an extensive topic in itself so only some of the initial comparisons are presented here. Although most of the established ionospheric models currently in use which are based on physics are not held at the Rutherford Appleton Laboratory, collaborations with RAL and the Universities have been established to produce these results.

The first model is an ionospheric ion-chemistry model developed by the University of Southampton. This model was developed to study the time varying effects of impulse events in the ionosphere, such as particle precipitation or aurora at high latitudes.

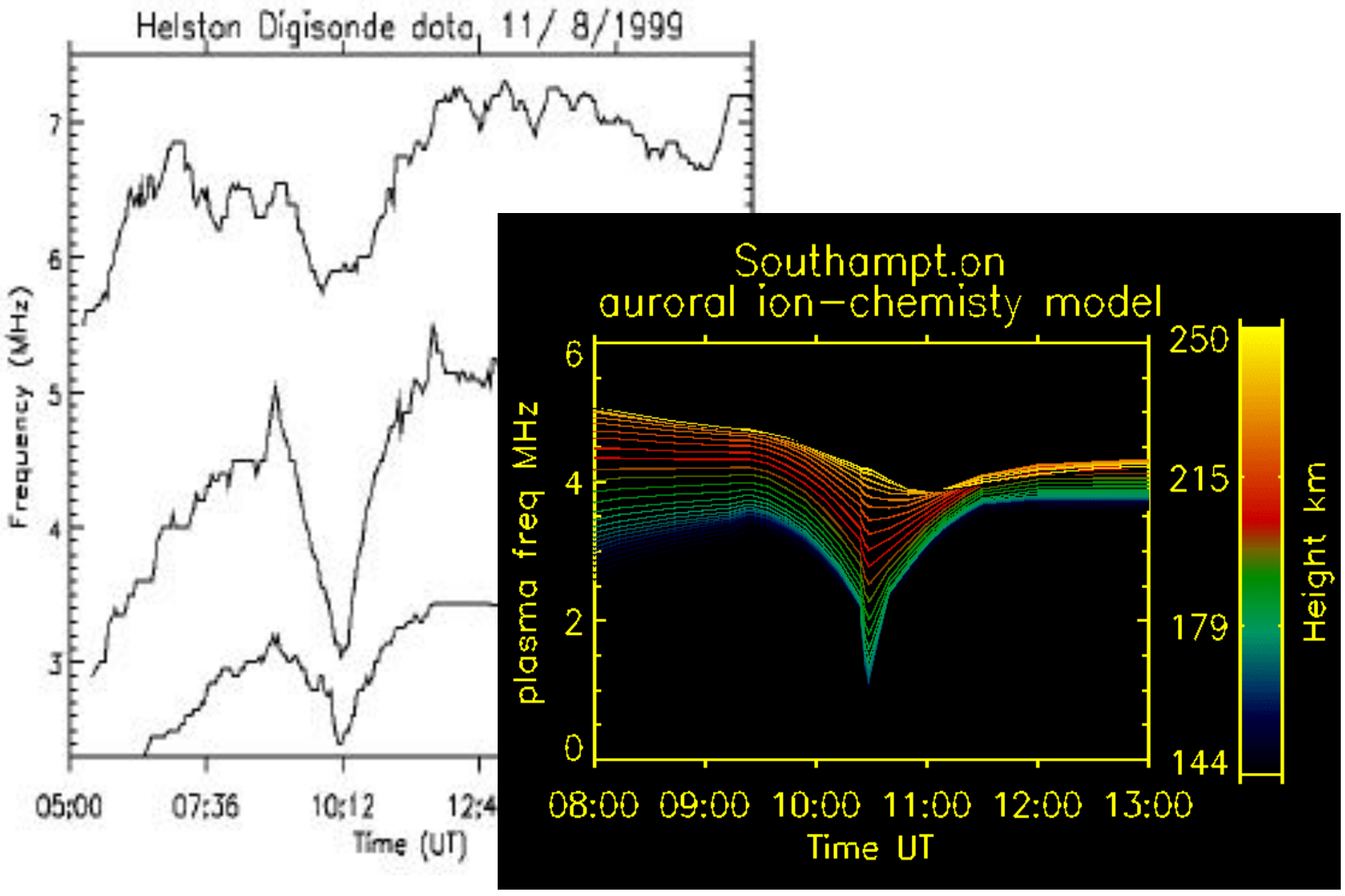

*Figure 30. Measurements of the ionospheric critical frequencies (left) and the initial simulation results from the Southampton ion-chemistry model (right). The colours represent the temporal response of different altitudes of the ionosphere.*

In Figure 30 the measurements of the changes in the ionospheric layers is shown along side the results of the ion-chemistry simulation. Although these are only the initial results it shows that the model is better at re-creating the lower altitude ionosphere than the upper ionosphere. The explanations for this are in the nature of the model and can be adapted. The model however has accurately predicted the step in gradient on the upward and downward leg before and after the maximum of the eclipse. Other results from this model have shown this to coincide with the collapse of the electron temperature. The temperature plays a pivotal part in determining the loss of electrons through effecting the particle collision rate. This model therefore would seem to be able to indicate the origins of the radio propagation results. However these results are provisional and more refinements are really needed for quantifiable comparisons.

The ion-chemisty model shown above is a 1D model and cannot show the geographical differences from the eclipse shadow as it moved across Europe. Another model has to be used here. In Figure 31 the University of London Global Thermosphere Ionosphere, Plamasphere model has produced a world wide map of the changes in the maximum frequency of the ionosphere due to the disturbance caused by the lunar shadow traversing the ionosphere at super sonic speeds.

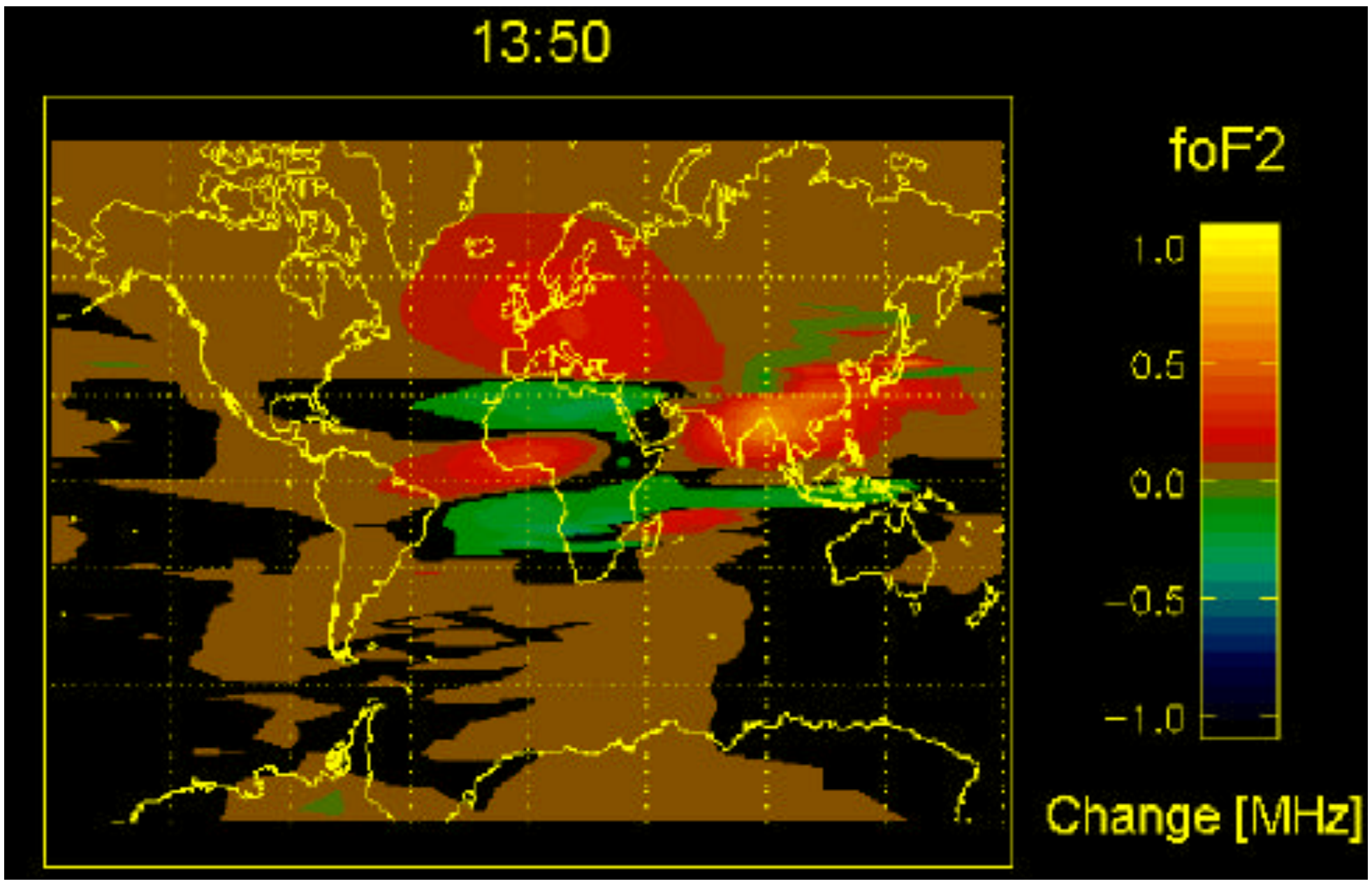

*Figure 31. The predictions of the effect of a global ionospheric model on the change of the maximum frequency of the F layer across the world as a consequence of the eclipse.( University of London).*

The use of the well described impulse disturbance of the eclipse for model validation clearly has implications for validating space weather predictions. Under no other circumstance can the cause of an observed effect be described so accurately.

# CONTACTS WITH THE MEDIA AND GENERAL PUBLIC

## THE ECLIPSE WEB SITE

One of them most successful outlets for contact with the general public for this event was through the Radio communications Agency's (RA) joint sponsorship for the main UK Eclipse website at www.eclipse.org.uk. The site was funded by Rutherford Appleton Laboratory (RAL), the Particle Physics and Astronomy Research Council (PPARC) (who also jointly funded the ionosonde project with RA), RA and the Royal Astronomical Society.

The web site was the major eclipse site for the 1999 solar eclipse in Britain. The front page of which is shown in Figure 32 with the link to the RA site. A short cut to the radio experiments was available at www.eclipse.org.uk/radio where there are detailed descriptions of how to do the medium wave general public and radio amateur experiments.

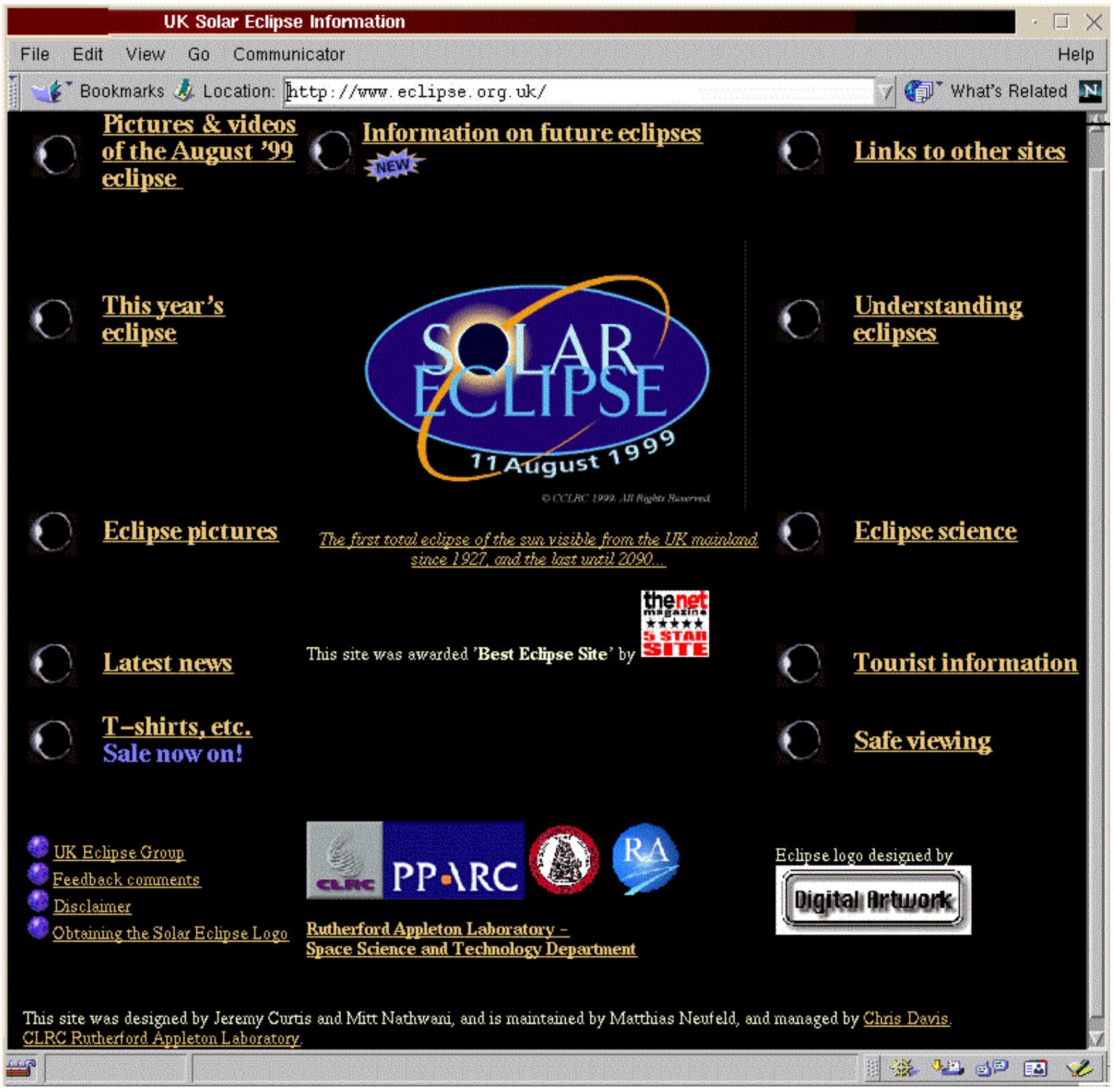

*Figure 32. The top page of the main UK Eclipse website. As one of the sponsors of the website the RA logo and link to RA webpages appeared on the top page.*

The web site was run on the World Data Centre Unix machines at Rutherford Appleton Laboratory. The plots in the two panels of Figure 33 illustrate the success of the website. The upper plot shows the total "hit rate" or number of people activating the website relative to the appearance of articles and interviews in the media. Although the website was a general website on the eclipse most of the media exposure was concerning the radio experiments. The lower plot shows the total number of web hits, which reached 16 million on the day of the eclipse.

# APPEARANCES OF THE RADIO EXPERIMENTS IN THE MEDIA

A lot of effort was directed to publicising the eclipse and the radio experiments during the eclipse. This effort was a joint effort and lead to the radio experiments appearing in 12 TV interviews, 15 radio broadcasts and 22 newspaper and periodical articles. A list of the specific items is shown in Table 3. Much of the success of the radio experiments with the media was associated with the recent discoveries made at RAL about the long term changes of the sun's corona and global warming and their possible confirmation using ionospheric measurements during eclipses [Davis *et al* 2000].

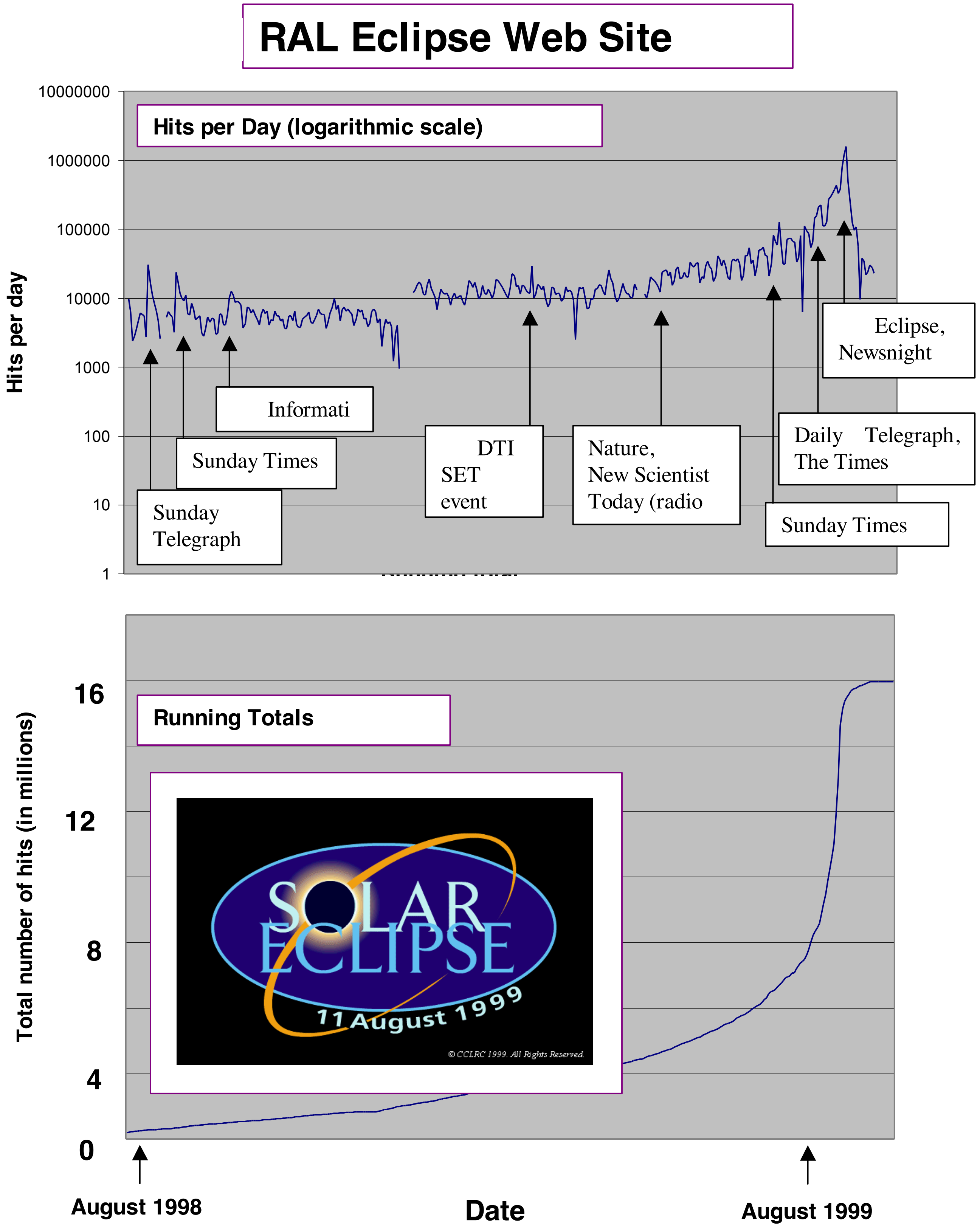

*Figure 33. The statistics for the eclipse web site. Top panel is the web hits per day relative to items on the radio experiments appearing in the media. The lower panel is the total number of web hits during the eclipse. (Plots thanks to Mattias Neufield).*

*Table 3. A LIST OF THE MEDIA ITEMS MADE DURING THE RADIO ECLIPSE PROJECT. Key: RAB: Dr. Ruth Bamford, CJD: Dr. Chris Davis, ML: Prof. Mike Lockwood*

| **TV Interviews** | **Who** |
|---|---|
| Sky News (Once at the lab prior to eclipse, once in Cornwall) | RAB/CJD |
| BBC News24 (and/or Breakfast News) | RAB/CJD |
| Newsnight (live interview by Chris Davis, lead item on 11th Aug) | CJD |
| 9 o'clock News (although may not have been used) | CJD |
| Channel 4 News | RAB |
| Yorkshire TV | CJD |
| Central News (twice) | RAB/ML |
| Sky at Night (twice) | CJD/RAB |
| NEWSPAPER ARTICLES (NATIONAL & INTERNATIONAL) | |
| The Daily Telegraph (four articles) | CJD/RAB |
| The Independent | CJD/ RAB |
| The Express | CJD |
| The Guardian | CJD |
| The Independent on Sunday | CJD/RAB |
| USA Today | CJD |
| Medium Wave News | CJD |
| Yorkshire Post | RAB |
| News for You (in Braille, the RNIB monthly journal)(twice) | RAB |
| **Newspapers (Local)** | |
| Oxford Mail | CJD/RAB |
| The Helston Packet | CJD |
| The West Country Gazette | CJD |
| **Magazine articles** | |
| New Scientist | CJD |
| Astronomy & Geophysics | RAB |
| New Electronics | RAB |
| RadCom | RAB |
| Radio Today | RAB |
| **Radio Interviews** | |
| Radio Sweden | CJD |
| Norwegian Radio | CJD |
| Thames Valley FM | CJD/RAB |
| Radio Cornwall (Twice) | RAB/CJD |
| Radio Wiltshire | RAB |
| Pirate FM (Twice) | CJD |
| World Service 'In the stars' | CJD |
| World Service (after the eclipse) | CJD |
| World Service (Brazilian section) | CJD |
| Radio 4 : The Today programme | RAB/CJD |
| Science programme | CJD |
| National Public Radio (NPR) | CJD |
| Australian National Radio | CJD |

# SUMMARY

Reported here are the details of the Radio Eclipse project conducted by the Rutherford Appleton Laboratory from 1998 to 2000. The event was a highly successful for public relations for radio with nearly 60 appearances on local and national TV, newspapers and periodicals. Nearly 1700 people responded to the general public medium wave experiment and 130 radio amateurs conducted experiments across Europe during the eclipse and 16 million people looked in on the general eclipse web site which included the radio experiments. Through these contacts many more than those who actively participated will have been introduced to aspects of radio reception that they were unaware of previously. The feedback from the general public and especially radio amateurs has been exceedingly positive.

The event of the eclipse was not only a highly successful public relations event it also has provided a large database of radio observations and ionospheric measurements.
The passage of the lunar shadow across the UK decreased the ionospheric critical frequencies by 20 to 35%. Sky wave radio propagation was effected from bands VLF to HF (3kHz to 30 MHz). Most notably the ionospheric absorption decreased by 10 to 40dBm which allowed some continental stations particularly on the medium wave to audibly fade in and fade out during the eclipse. The level of the effect on the signal strength was quite comparable with the drop during night-time as predicted from the ITU observations.

In order to limit the length of this report, many aspects of the eclipse project have been excluded by necessity. These include the results from the Total Electron Content (TEC) across Europe, oblique ionospheric sounding between the UK and France, the first ever observations of the wind changes at ionospheric altitude recorded by the meteor scatter radar nor the mapping of the ionospheric disturbance in 4D using the COST 251 ionosonde network and the comparisons of these maps with ionospheric models. Nor is there a description of the work on global warming and the changes in the solar corona as deduced by Helston ionospheric measurements that were made during the eclipse. This work has been omitted because the topic is not directly relevant to radio propagation however the media did generally run the two aspects of the ionospheric measurements together and each benefited from the association.

The large and unique database collected on the many aspects of the eclipse disturbance can be used to test ionospheric prediction models. This is only possible because the disturbance of the eclipse can be exactly characterised and now the predictions and observations can compared.

In conclusion, all of the main objectives of the project were met. That is:

(i) Public awareness of the effects of sun and night-time conditions on radio reception in both the medium and short wave bands was increased.

(ii) Better links between propagation researchers and the short wave radio user community by working together on a specific event were established.

(iii) The unique circumstance of a total solar eclipse has been used to examine ionospheric models.